\documentclass[prb,twocolumn,floatfix,showpacs]{revtex4}
\usepackage{psfrag,epsfig} 
\usepackage{graphicx}
\usepackage{amssymb}
\usepackage{amsmath}

\begin{document}
\title{Relaxation dynamics of an exactly solvable electron-phonon model}

\author{D.M.~Kennes} \affiliation{Institut f\"ur Theoretische Physik A and JARA--Fundamentals of 
Future Information Technology, RWTH Aachen University, 52056 Aachen, Germany}  
\author{V.~Meden} \affiliation{Institut f\"ur Theoretische Physik A and JARA--Fundamentals of 
Future Information Technology, RWTH Aachen University, 52056 Aachen, Germany} 

\begin{abstract}
We address the question whether observables of an exactly solvable model of electrons coupled to 
(optical) phonons relax into large time stationary state values and investigate if the asymptotic 
expectation values can be computed using a stationary density matrix. Two initial nonequilibrium 
situations are considered. A sudden quench of the electron-phonon coupling, starting from the 
noninteracting canonical equilibrium at temperature $T$ in the electron as well as in the phonon 
subsystems, leads to a rather simple dynamics. A richer time evolution emerges if the initial state 
is taken as the product of the phonon vacuum and the filled Fermi sea supplemented by a 
highly excited additional electron. Our model has a natural set of constants of motion, with 
as many elements as degrees of freedom. In accordance with earlier 
studies of such type of models we find that expectation 
values which become stationary can be described by the density matrix of a generalized Gibbs 
ensemble which differs from that of a canonical ensemble. 
For the model at hand it appears to be evident that the eigenmode occupancy 
operators should be used in the construction of the stationary density matrix. 
\end{abstract}

\maketitle

\section{Introduction}
\label{sec:intro}

It is of fundamental interest to reveal the conditions under which 
an isolated quantum system for $t \to \infty$ relaxes into a 
state  that can be described by a stationary density matrix. 
Furthermore, a detailed understanding 
of the relaxation process is desirable. 
We here characterize relaxation by considering the time 
evolution of expectation values of observables.
Starting out from a nonequilibrium situation at $t=0$ described 
by the initial density matrix $\hat \rho_{\rm i}$ we ask whether the 
expectation value 
\begin{eqnarray}
\label{expec}
\left< \hat O_{\rm l} \right>_{\hat  \rho(t) }= \mbox{Tr} \left[\hat  \rho(t) 
 \hat O_{\rm l} \right]
\end{eqnarray}
of a {\it local observable} $ \hat O_{\rm l}$ for time 
$t \to \infty$ approaches a constant value which can also be computed considering the 
stationary density matrix $\hat \rho_{\rm st}$
\begin{eqnarray}
\label{expec_limit}
\lim_{t \to \infty} \left< \hat O_{\rm l} \right>_{\hat \rho(t)} = 
\mbox{Tr} \left[ \hat \rho_{\rm st}
 \hat O_{\rm l}\right] \; .
\end{eqnarray}
Here $ \hat \rho(t)$ denotes the statistical operator at time $t$ which follows from 
solving the von Neumann equation for the given initial condition $\hat \rho(t=0)=
\hat \rho_{\rm i}$. A ``local observable'' is defined as one 
which only contains degrees of freedom from a subsystem ${\mathcal S}$ of the isolated quantum 
system $ {\mathcal Q}$. We focus on measurements in subsystems and one must thus 
be careful in interpreting $\hat \rho_{\rm st}$ as the density matrix describing the 
stationary state of the entire quantum system ${\mathcal Q}$. 
For a meaningful description of the stationary state $\hat \rho_{\rm st}$ should 
be independent of the chosen observable.
To avoid recurrence effects one 
has to perform the thermodynamic limit of $\mathcal Q$ which is often done by taking the 
thermodynamic limit $V_{\mathcal E} \to \infty$, with the volume $V_{\mathcal E}$ of the {\it environment} 
${\mathcal E} = {\mathcal Q}/{\mathcal S}$, keeping $V_{\mathcal S}$ fixed. 

Relaxation to a time independent 
expectation value in the strict sense can only occur after the thermodynamic limit  
has been taken. Alternatively one can address the question whether a ``quasi 
stationary state'' is reached in a {\it finite} system.
By this we understand a situation in which expectation values ``fluctuate'' around 
a constant value which can be extracted by averaging over time. If such a state is reached one can 
ask if the time averaged value can be computed using a stationary statistical ensemble.
In the averaging it might even be meaningful to increase the time interval beyond the characteristic 
time $t_{\rm r} = L/v$---denoted recurrence time in what follows---with $L$ being a typical length and $v$ 
a typical velocity of the system (see below).

In {\it equilibrium} statistical physics we commonly work with {\it thermal} ensembles as the 
ones describing the state. They are characterized by the density matrix 
\begin{eqnarray}
\label{candensitymatrix}
\hat \rho = \frac{1}{Z} \, e^{- \sum_{j=1}^n \eta_j \hat I_j } \; , \;\;\; 
Z = \mbox{Tr} \, 
e^{- \sum_{j=1}^n \eta_j \hat I_j } \; ,
\end{eqnarray}
with the partition function $Z(\{ \eta_j \})$. 
The sum usually runs over only a few terms containing operators $\hat I_j$, such as the 
Hamiltonian $\hat H$ and the particle number operator $\hat N$ corresponding to the macroscopic variables energy and 
particle number. 
The Lagrange multipliers $\eta_j$ are fixed such that the expectation values of the $\hat I_j $ 
take given values $I_j^{(0)}$ (e.g. given average energy and particle number)
\begin{eqnarray*}
\left<  \hat I_j \right>_{\hat \rho} = I_j^{(0)} \; .
\end{eqnarray*}
Chosing the corresponding Lagrange parameters in Eq.\ (\ref{candensitymatrix}) 
maximizes the entropy $S=\mbox{Tr} \, [ \hat \rho \ln{\hat \rho^{-1}} ]$ under the constraint of 
fixed $I_j^{(0)}$.\cite{Jaynes}  
Within the observables of a closed system (with fixed particle number) the Hamiltonian plays a special 
role as in many situations the expectation value of the energy is the only conserved 
quantity. The corresponding Lagrange multiplier 
is the inverse temperature  $\beta=1/T$ and the thermal ensemble with only the energy expectation value fixed 
is the {\it canonical} one.  

Jaynes\cite{Jaynes} studied generalized ensembles---now commonly referred 
to as generalized Gibbs ensembles (GGEs)---in which additional observables $\hat I_j$ 
besides the energy are assumed to take a given expectation 
value (fixed $I_j^{(0)} $) and together with the corresponding Lagrange multipliers enter the sum in 
Eq.\ (\ref{candensitymatrix}). 

Starting out with the initial nonequilibrium state given by 
$\hat \rho_{\rm i}$ and under the assumption that  
$\left< \hat O_{\rm l} \right>_{\hat \rho(t)}$ converges for 
$t \to \infty$ one might expect that the stationary expectation value can be 
computed from Eq.\ (\ref{expec_limit}) 
with $ \hat \rho_{\rm st} = \hat \rho_{\rm can} $ and 
\begin{eqnarray}
\label{canny} 
\hat \rho_{\rm can} =  e^{- \beta \hat H} / Z_{\rm can} \; .  
\end{eqnarray} 
This is known as {\it thermalization.}
The inverse temperature $\beta$ is set by the constraint 
\begin{eqnarray*} 
\left< \hat H \right>_{\hat \rho_{\rm can}} = \left< \hat H \right>_{\hat \rho_{\rm i}} \; .
\end{eqnarray*}  
In particular, this is the expected behavior if the energy is the {\it only} (independent) 
constant of motion.  

Recent experiments in the field of ultracold atoms\cite{coldex} led to a revived interest into 
relaxation dynamics. Due to the long coherence time such systems are ideal candidates to study the 
relaxation into a stationary state in a controlled setup. In the experiments an equilibrium 
state is disturbed by a sudden {\it quench} of system parameters. The experiments led to several 
interesting theoretical studies in which models, usually considered in the field of quantum many-particle 
physics, were investigated concerning their relaxation properties. In analogy to the experiments in 
most studies the system is assumed to be in an {\it eigenstate} (e.g. the noninteracting 
groundstate) or a {\it canonical thermal state} with fixed temperature $T$ of the {\it initial} 
many-body Hamiltonian $\hat H_{\rm i}$. At $t=0$ model parameters, in most cases the 
two-particle interaction being a crucial element of the models, is quenched instantaneously 
to a different value and the dynamics of the initial state under 
the time evolution given by the final Hamiltonian $\hat H_{\rm f}$ is computed. 

The investigations can be grouped in three classes. Analytical studies of models 
which can be solved 
exactly,\cite{Rigol1,Cazalilla,Perfetto,Rigol2,Cramer1,Gangardt,Barthel,Eckstein,Kollar} 
numerical studies,\cite{Kollath,Manmana,Cramer2,Flesch,Barmettler,Rigol3} and approximate analytical 
studies.\cite{Moeckel,Hackl} For most of the considered models the relaxation   
properties of a restricted set of observables was computed. 
A more general perspective on the problem for a certain class of models is taken in 
Refs.~\onlinecite{Calabrese} and \onlinecite{Fioretto} using methods of boundary critical phenomena and conformal 
field theory. The ultimate goal of the 
``case studies''  is (i) to derive criteria which a priori allow to answer the question whether 
certain (local) observables become stationary and (ii) to construct the stationary density 
matrix $\hat \rho_{\rm st}$, which is the appropriate ensemble by which the asymptotic 
expectation values can be computed. 
It is of particular interest to understand the conditions under which the latter becomes
the canonical one and the system thermalizes. From our considerations it is plausible to expect  
that the number and character of the constants of motion of a specific model are of crucial importance 
in answering the questions addressed above. Roughly speaking, if the number 
of constants of motion is large the time evolved state contains a lot of information 
from the initial state and thermalization cannot be expected.  
 In all the studies mentioned above single component systems containing either 
bosons or fermions were studied. Besides the asymptotic long time behavior transient 
nonequilibrium effects such as ``collapse and revival'' were investigated. 

From the analytical\cite{Rigol2,Cazalilla,Cramer1,Barthel,Eckstein,Kollar,Calabrese}   
and numerical\cite{Manmana,Rigol3} studies of exactly solvable 
models increasing evidence was collected 
that {\it if} the expectation value of an observable $\hat O$ approaches a constant large time 
limit, the latter cannot be obtained using the thermal canonical density matrix.  
Instead stationary density matrices of the {\it GGE type} Eq.\ (\ref{candensitymatrix}), 
corresponding to situations with more restrictions (than a fixed average energy) set by 
the initial state having a density matrix 
\begin{eqnarray}
\label{GGEdensitymatrix}
\hat \rho_{\rm GGE} = \frac{1}{Z_{\rm GGE}} \, e^{- \sum_{j=1}^n \eta_j \hat I_j } \; , \;\; Z_{\rm GGE}= 
\mbox{Tr} \, e^{- \sum_{j=1}^n \eta_j \hat I_j } 
\end{eqnarray}
turned out to be promising candidates. On the basis of the above considerations this is 
not surprising,  
as the models considered are characterized by more integrals of motion than just 
the energy and  
one thus expects for the stationary state an increased ``memory'' of the initial state. 
A set of constants of motion were taken as the $\hat I_j$ and the Lagrange multipliers 
$\eta_j$ were determined such that
\begin{eqnarray}
\label{require}
 \left<  \hat I_j \right>_{\hat \rho_{\rm i}} =   \left<  \hat I_j 
\right>_{ \hat \rho_{\rm GGE}} \; .
\end{eqnarray}
The choice of conserved observables is not unique. E.g. with $\hat H$ being a constant of motion
the same holds for $\hat H^2$, and it was shown\cite{Kollar} that under certain conditions 
the GGE expectation value of a given observable $\hat O$ 
\begin{eqnarray}
\label{cons}
\left< \hat O\right>_{\hat \rho_{\rm GGE}} = \mbox{Tr} \,  
\left[ \hat \rho_{\rm GGE} \hat O \right]
\end{eqnarray}
might depend on the selected set. Thus the GGE describes the stationary state only if 
the ``correct'' set of ``independent'' $\hat I_j$ is chosen. For a general model 
with several possible sets of constants of motion up to now it is not clear how to select 
the ``correct'' set a priori. 

In cases in which the system's Hamiltonian can be brought into the form 
\begin{eqnarray}
\label{canham}
\hat H = \sum_{k} \lambda(k) \hat \psi^\dag_k \hat \psi_k  \; ,
\end{eqnarray}
with either fermionic or bosonic creation and annihilation operators $\psi_k^{(\dag)}$
and quantum number $k$ the eigenmode occupation operators   $\hat n_k = \hat \psi^\dag_k \hat 
\psi_k $ constitute a natural set of conserved quantities, with as many elements as degrees of 
freedom. Then the statistical expectation value taken with the $\rho_{\rm GGE}$ set up with 
$\hat I_k= \hat n_k$ and Lagrange multipliers fixed by the constraints  Eq.\ (\ref{require}) 
led for generic model parameters to the correct large time limit of (certain) 
observables {\it provided} the latter exists\cite{Cazalilla,Barthel,Kollar} 
and the initial states are homogeneous.\cite{Aditi} 

Cold atom gases is not the only subfield of modern condensed matter physics in which 
understanding the relaxation dynamics is of crucial importance and quenches are not the 
only process leading to an interesting time evolution. In the area of photoexcited semiconductors much 
effort has been put into measuring (pump-probe techniques) and understanding the relaxation 
dynamics of highly excited electrons coupled to optical phonons; for a recent review see 
Ref.\ \onlinecite{Kuhn}. The long-time asymptotics of the electron-phonon system is difficult to study 
experimentally due to the strong coupling to other degrees of freedom (e.g. acoustic phonons and holes), 
but the short-time (transient) dynamics shows interesting non-Markovian effects requiring a treatment 
beyond the use of Boltzmann equations.\cite{Kuhn,Leitensdorfer,VM1,VM2} 
A detailed understanding of relaxation also plays a major role in the field of 
condensed matter system based quantum information processing.\cite{QI} Here the strong 
coupling of the degrees of freedom envisaged as quantum-bits to the environment usually 
leads to a coherence time to short to perform substantial information processing. Gaining insights 
into the relaxation process might lead to ways to circumvent this obstacle and significantly increase
the coherence time.

We here supplement the recent ``case studies'' on the relaxation dynamics of either bosonic or fermionic 
correlated systems by analytically investigating the time evolution of a {\it two-component} model of 
electrons coupled to phonons. The model naturally contains two subsystems---the electron and the 
phonon systems---and 
an observable can be considered as {\it local} if it contains only fermionic or bosonic degrees
of freedom.  
We first 
consider the time evolution resulting from a quench of the electron-phonon coupling from zero to a finite value
starting with the noninteracting canonical equilibrium at temperature $T$ in the electron as well as the phonon 
subsystems. In 
{\it addition} we study the dynamics inferred by the interacting Hamiltonian out of a pure state given by 
the product of the phonon vacuum and the filled Fermi sea supplemented by a highly excited additional 
electron of momentum $k_0$.
The {\it short-time} dynamics of our model starting with this initial state was earlier discussed in 
the context of optically excited semiconductors\cite{VM1} and used to explain results of pump-probe 
experiments.\cite{Leitensdorfer,VM2} While the time evolution of observables, 
in particular the electron or phonon momentum distribution function and the subsystem energies, is rather 
simple for the quench a rich dynamics is found in case of the ``$k_0$-excitation''. We show that in 
the large time limit $t \to \infty$ the subsystem energies (and the energy in the 
electron-phonon coupling) converge to stationary values for both nonequilibrium initial states. The same 
holds for the electron momentum distribution, while for the 
phonon momentum distribution function convergence is only achieved after averaging over a small momentum 
interval. Our model can be brought into the form Eq.\ (\ref{canham}). It thus 
contains (at least) as many constants of motion as degrees of freedom.  In accordance with earlier 
studies\cite{Cazalilla,Barthel,Kollar} we find that for both initial states the expectation 
values of observables which become stationary  can be described by the density matrix of a 
generalized Gibbs ensemble with the eigenmode occupation operators chosen as the $\hat I_j$. 

The rest of this paper is organized as follows. In Sect. \ref{sec:model} we introduce our 
one-dimensional (1d) electron-phonon model and specify our initial states. The bosonization 
of the fermionic field operator, which allows to obtain analytical results for the time evolution of 
fermionic observables, is discussed in Sect. \ref{sec:method}. 
Section \ref{sec:GGE} is devoted to the GGE for the model at hand. In Sects. \ref{sec:quench} and 
\ref{sec:k_0_ex} we discuss the relaxation dynamics for our two distinct initial states and compare
the long-time expectation values to those obtained from the GGE. Finally, our results are summarized in 
Sect. \ref{sec:summary}.

\section{The electron-phonon model and the nonequilibrium  initial states}
\label{sec:model}

We consider a model of electrons on a 1d ring of length $L$ (periodic boundary conditions) 
coupled to phonons by a Holstein type electron-phonon interaction given by the Hamiltonian 
\begin{equation} 
\label{hamiltonian1}
\hat H=\hat H_{\rm e}+\hat H_{\rm p}+\hat H_{\rm ep} \; ,
\end{equation}
with
\begin{eqnarray}
\label{ham1}
\hat H_{\rm e} &=&\sum_k  \epsilon_k \left(\hat a_k^{\dagger} \hat a_k^{}
-\left< \hat a_k^{\dagger} \hat a_k^{} \right>_0 \right)  \; , \\
\label{ham2}
\hat H_{\rm p} & = & \sum_{q>0} \omega_q \hat B_q^{\dagger} \hat B_q^{} \; , \\
\label{ham3}
\hat H_{\rm ep} & = & \left( \frac{2 \pi}{L} \right)^{1/2}  
\sum_{{k \atop q>0}} g(q)
\left( \hat a_{k+q}^{\dagger} \hat a_{k}^{} \hat B_{q}^{} + \mbox{H.c.} \, \right) \; .             
\end{eqnarray}
Here $\hat a_k$ ($\hat B_q$) is the annihilation operator of an
electron (phonon) with momentum $k$ ($q$) and $\left<...\right>_0$ is the expectation value in the
noninteracting groundstate (normal ordering). The momentum dependent electron-phonon coupling is 
denoted by $g(q)$. To be specific we consider the (simple) form $g(q)=g \Theta(q_{\rm c}-q)$. 
We restrict our considerations to a single branch of chiral (right-moving)
spinless fermions with linear single-particle dispersion $\epsilon_k=v_{\rm F}(k-k_{\rm F})$, the 
Fermi velocity $v_{\rm F}$, and the Fermi momentum $k_{\rm F}$ which, without loss of generality, is set to zero 
in the following ($k_{\rm F}=0$). 
We assume that the fermion states do not have a lower bound (Dirac model) and all momentum 
states with $k<0$ are filled in the groundstate (filled Fermi sea). For the sake of convenience the
Fermi momentum $k_{\rm F}$ corresponds to the first empty instead of the last 
occupied state. 
Divergencies possibly resulting
from these states are regularized by the normal ordering.   
Furthermore, we focus on optical phonons (Einstein model) with a single 
energy $\omega_q=\omega_0$. 

The assumed $q$-dependences of the phonon dispersion and the
electron-phonon coupling can be relaxed without spoiling the possibility of an exact analytical 
solution of the model. Varying the function $g(q)$ only leads to minor changes in the short-time 
dynamics\cite{VM1} (as long as $g(0)$ remains finite) and we do not see any physical reason why 
our main results for the {\it long-time} relaxation obtained here should dependent on the precise form of 
$g(q)$. Our model is a variant  of the purely fermionic Tomonaga-Luttinger 
model\cite{tl,Haldane}---more precisely what is called the chiral $g_4$-model\cite{Solyom}---and similar 
to this case the linear fermion dispersion is crucial for an exact analytic solution. 
Equilibrium properties of the above model, also considering acoustic instead of optical phonons,
were discussed earlier.\cite{Wentzel,Engelsberg,Apostol,VM3,VM4} The quench dynamics of the 
Tomonaga-Luttinger model was studied in Ref.\ \onlinecite{Cazalilla}.

To diagonalize our Hamiltonian Eqs.\ (\ref{hamiltonian1})-(\ref{ham3}) we first introduce 
the electron density operator
\begin{equation}
\label{density}
\hat d_q=\sum_k \hat a^{\dagger}_k \hat a^{}_{k+q} \;\; ,
\end{equation}
with $q > 0$. For $q=0$ we define the electron number operator relative
to the groundstate (filled Fermi sea) as
\begin{equation}
\label{particlenumber}
\hat{N}= \sum_k \left( \hat a^{\dagger}_k \hat a^{}_k- \left< \hat a^{\dagger}_k \hat a^{}_k
\right>_0 \right) \;\; .
\end{equation}
With a proper normalization the densities $\hat d_q$ obey Bose commutation relations.\cite{Haldane}
If one defines 
\begin{equation}
\label{bosons}
\hat b_q^{} = \left( \frac{2\pi}{qL} \right)^{1/2} \hat d_q \;\;\;\; , \;\;\;\;
\hat b_q^{\dagger} = \left( \frac{2\pi}{qL} \right)^{1/2} \hat d_{-q}  \; ,
\end{equation}
for $q>0$, the commutation relations read
\begin{equation}
\label{commutator}
\left[ \hat b^{}_q, \hat b^{\dagger}_{q'}  \right] = \delta_{q,q'} 
\;\;\;\; , \;\;\;\;
\left[ \hat b^{}_q, \hat b^{}_{q'}  \right] = 0 \; .
\end{equation}
The electron-hole excitations in
$H_{\rm ep}$ can straightforwardly be written linearly in the boson operators 
$b^{}_q$ and $b^{\dagger}_q$. For 1d systems with a linear
dispersion it is in addition possible to write the kinetic energy of the
fermions as \cite{Kronig,Haldane}
\begin{equation}
\label{kronig}
H_{\rm e}=\sum_{q>0} \varepsilon_q b_q^{\dagger} b_q^{} + c(\hat{N}) \; ,
\end{equation}
where $c(\hat{N})$ only contains the fermionic particle number operator.
In the following this term can be droped because we are working
in a sector of the Hilbert space with constant particle number. 
We can now formulate  Eqs.\ (\ref{hamiltonian1})-(\ref{ham3}) in terms of the bosonic fermion
densities and the phonons as
\begin{eqnarray}
\label{hambos}
\hat H & = &  \sum_{q>0} v_{\rm F} q \hat b_q^{\dagger} \hat b_q^{} + \omega_0 \sum_{q>0}  \hat B_q^{\dagger}\hat B_q^{} \nonumber \\ 
&& + g \sum_{0 < q \leq q_{\rm c}} \sqrt{q} \left( \hat b_q^{\dagger} \hat B_q + \mbox{H.c.} \right) \; . 
\end{eqnarray}
From this expression it becomes evident that the groundstate of the
model is still the tensor product of the Fermi sea---corresponding to the 
vacuum of the bosonic fermion density $\hat b_q$---and the phonon vacuum.
Therefore all groundstate expectation values, like e.g.\ the fermionic momentum
distribution, are given by the noninteracting ones. 
 
\begin{figure}[tb]
\includegraphics[width=.8\linewidth,clip]{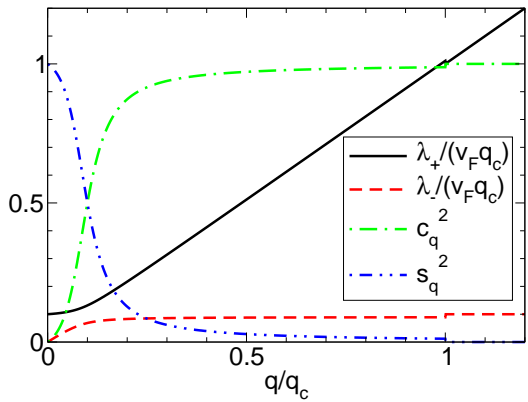}
\caption{(Color online) Eigenmodes $\lambda_\pm(q)$ and coefficients $c_q^2$ and $s_q^2$ 
of the eigenvectors for  
$\Gamma=0.01$ and $\Omega=0.1$. Note the sharp transition (discontinuity) to the noninteracting 
values at $q_c$.}
\label{fig1}
\end{figure}

Using a canonical transformation the problem of the coupled bosonic modes
Eq.\ (\ref{hambos}) can be brought in the form of 
Eq.\ (\ref{canham}).\cite{Wentzel} The transformation is given by 
\begin{eqnarray}
\hat b_q & = & \hat \alpha_q c_q - \hat \beta_q s_q \; ,  \nonumber \\
\label{trafo}
\hat B_q & = & \hat \alpha_q s_q + \hat \beta_q c_q \; ,
\end{eqnarray}
with 
\begin{eqnarray}
\label{trafokoeff}
c_q^2 = \frac{\left| \lambda_+(q) - \omega_0 \right|}{ \lambda_+(q) -  \lambda_-(q)} \; ,
\; \; \;  s_q^2 = \frac{\left| \lambda_-(q) - \omega_0 \right|}{ \lambda_+(q) -  \lambda_-(q)} 
\end{eqnarray}
and the mode energies 
\begin{equation}
\label{energies}
\lambda_{\pm}(q) = \frac{1}{2} \left\{ v_{\rm F} q +\omega_0 \pm
\sqrt{\left(v_{\rm F} q  -\omega_0 \right)^{2} +4 g^{2} 
q  \Theta(q_{\rm c}-q) } \right\}  \; .
\end{equation}
Note that $c_q^2+s_q^2=1$ for all $q>0$ and that $c^2_q=1$, $s^2_q=0$ for $q>q_{\rm c}$. 
Here we focus on the case $v_{\rm F} q_{\rm c} > \omega_0$. 
In the new bosonic operators $\hat \alpha_q$ and $\hat \beta_q$ the Hamiltonian
reads
\begin{equation}
\label{newhamiltonian}
\hat H=\sum_{q>0} \left[ \lambda_+(q) \hat \alpha_q^{\dagger} \hat \alpha_q^{} +
\lambda_-(q) \hat \beta_q^{\dagger} \hat \beta_q^{} \right] \; .
\end{equation}
In order to obtain a stable groundstate the boson energies
$\lambda_{\pm}(q)$ have to be larger than zero.\cite{Wentzel}
For a given $\omega_0$ this leads to a restriction of the coupling strength 
$g$ that can be used. The dimensionless parameters of the model are 
$ \Gamma = \frac{g^2}{v_{\rm F}q_{\rm c}}$ for the electron-phonon coupling, $\Omega = 
\frac{\omega_0}{v_{\rm F} q_{\rm c}}$ for the phonon frequency, and $\nu = \frac{2 \pi}{L q_{\rm c}}$ for the 
inverse of the ring length. Stability requires that $\Gamma < \Omega$. The momentum dependence of the 
eigenmode energies $\lambda_{\pm}(q)$ and the coefficients $c_q^2$ and $s_q^2$ are shown in Fig.\ \ref{fig1}
for a typical set of parameters with $\Gamma=0.01$ and $\Omega=0.1$. 

With Eqs.\ (\ref{newhamiltonian}) and (\ref{trafokoeff}) 
computing the dynamics of the 
phonon ladder operators and fermionic densities becomes simple. 
We find  
\begin{eqnarray}
\label{Bqt}
\hat B_q(t) & = & c_q s_q  \left( e^{-i\lambda_+(q)t} - e^{-i\lambda_-(q)t}
\right) \hat b_q  \nonumber \\
&& +    \left( s_q^2 e^{-i\lambda_+(q)t} + c_q^2 e^{-i\lambda_-(q)t}
\right) \hat B_q  
\end{eqnarray}
and 
\begin{eqnarray}
\label{bqt}
\hat b_q(t) & = &  \left( c_q^2 e^{-i\lambda_+(q)t} + s_q^2 e^{-i\lambda_-(q)t}
\right)  \hat b_q \nonumber \\
&& + c_q s_q  \left( e^{-i\lambda_+(q)t} - e^{-i\lambda_-(q)t}
\right) \hat B_q \; .
\end{eqnarray}
The time dependence of expectation values
of observables which can be written in terms of the $B_q^{(\dag)}$ and 
$b_q^{(\dag)}$ can thus be expressed by expectation values 
taken with the initial density matrix $\hat \rho_{\rm i}$. 
This gives us direct access to the dynamics of the phonon momentum distribution 
function 
\begin{eqnarray}
\label{phodisdef}
N(q,t) = \left< \hat B_q^\dag(t) \hat B_q(t) \right>_{\hat \rho_{\rm i}} =   
\left<  \hat B_q^\dag(0) \hat B_q(0) \right>_{\hat \rho(t)}
\end{eqnarray}
and that of the subsystem energies for both nonequilibrium states considered. 
Alternatively with Eqs.\ (\ref{Bqt}) and (\ref{bqt}) the time evolution of the density 
matrix can be given in a closed form if $\hat \rho_{\rm i}$ can be expressed in terms of the 
$b_q^{(\dag)}$ and $B_q^{(\dag)}$ (see below). 
To compute the fermionic momentum distribution function 
\begin{eqnarray*}
n(k,t) = \left< \hat a_k^\dag(t) a_k(t) \right>_{\hat \rho_{\rm i}} 
= \left< \hat a_k^\dag(0) a_k(0) \right>_{\hat \rho(t)}
\end{eqnarray*}
one  has to establish a relation between the fermionic operators 
$a_{k}^{(\dag)}$ and the $b_q^{(\dag)}$. This {\it bosonization} (of the 
field operator) will be discussed in Sect.\ \ref{sec:method}.

We consider two different initial situations at time $t=0$. In the first one the 
{\it decoupled} ($g=0$) electron-phonon system is initially  assumed to 
be in a thermal state with a common temperature $T=1/\beta$ of the electron 
and phonon subsystems corresponding to the noninteracting canonical 
ensemble. It is determined by the initial density matrix (superscript q for ``quench'') 
\begin{eqnarray} 
\label{iniquench}
\hat \rho_{\rm i}^{\rm q} = e^{- \hat \beta H_{\rm e}} \otimes e^{-\hat \beta H_{\rm p}} / Z \; .
\end{eqnarray}
The second initial density matrix is given by a {\it pure state} (superscript $k_0$ for 
$k_0$-excitation) 
\begin{eqnarray}
\label{inik_0}
\hat \rho_{\rm i}^{k_0} = \left| \Psi_{\rm i} \right> \left< \Psi_{\rm i} \right|
\end{eqnarray}
with 
\begin{eqnarray}
\label{inistate}
\left| \Psi_{\rm i} \right> = \hat a_{k_0}^\dag \left| \mbox{FS} \, \right> \otimes \left| \mbox{vac} 
\right> \; .
\end{eqnarray}
Here $ \left| \mbox{FS} \, \right>$ denotes the filled Fermi sea (vacuum with respect to the $\hat b_q$) 
and $ \left| \mbox{vac} \right> $ the 
phonon vacuum. One can think of this state being (approximately) realized in a doped semiconductor (Fermi sea in conduction 
band) in which an additional ``hot'' electron with momentum $k_0$ is optically pumped from the valence band 
into the conduction band. Both initial conditions correspond to nonequilibrium states if the time evolution is 
performed with the Hamiltonian Eq.\ (\ref{hambos}) for $g \neq 0$. We note in passing that starting out with the 
noninteracting groundstate ($T=0$) would {\it not} lead to a time dependence of expectation values after a sudden 
quench of $g$ as in the present model this state remains the groundstate even for $g \neq 0$.

\section{Bosonization of the fermions}
\label{sec:method}

To
calculate fermionic expectation values like the momentum 
distribution function  we use the bosonization of the fermionic 
field operator
\begin{equation}
\label{field}
\hat \psi^{\dagger}(x)= \frac{1}{\sqrt{L}} \sum_{k} e^{-ikx} \hat a_k^{\dagger} \;\; .
\end{equation}
One can prove the operator identity \cite{Haldane}
\begin{equation}
\label{bosonization1}
\hat \psi^{\dagger}(x)= \frac{e^{- i x\pi/L}}{\sqrt{L}} e^{-i
\hat \Phi^{\dagger}(x)} \hat U^{\dagger} e^{-i \hat \Phi(x)} \;\; ,
\end{equation}
with
\begin{equation}
\label{bosonization2}
\hat \Phi(x)=\frac{\pi}{L} \hat{N} x -i \sum_{q>0} e^{iqx} \left(
\frac{2\pi}{L q} \right)^{1/2} \hat b_q \;\; ,
\end{equation}
where $\hat U^{\dagger}$ denotes a unitary fermionic raising operator 
which commutes with the $\hat b_q^{(\dag)}$ and maps
the $N$-electron groundstate to the $(N+1)$-electron groundstate. 
As we are interested in the dynamics for a fixed particle number neither 
the term proportional to $\hat N$ in Eq.\ (\ref{bosonization2}) nor 
the fermionic raising operator $\hat U^\dag$ affect the result and both can be 
dropped in the following. Using Eqs.\ (\ref{bosonization2}), 
(\ref{bosonization1}) and the back transform 
\begin{eqnarray*}
\hat a_k^\dag = \frac{1}{\sqrt{L}} \int_{-L/2}^{L/2} e^{i k x } \psi^\dag(x) \, dx 
\end{eqnarray*}
of Eq.\ (\ref{field}) as well as Eq.\ 
(\ref{bqt}) the time dependence of $ a_k^{(\dagger)}$ can be expressed in terms of the $b_q^{(\dag)}$ 
and $B_q^{(\dag)}$. As further deepened below this allows us to give a closed expression for the time 
evolution of the fermionic momentum distribution $n(k,t)$ as well as its value in the 
appropriate GGE.   

\section{The generalized Gibbs ensemble}
\label{sec:GGE}

As shown in Sect. \ref{sec:model} our coupled electron-phonon Hamiltonian can be brought into the 
diagonal form 
Eq.\ (\ref{canham}) with the bosonic operators $\hat \alpha_q^{(\dag)}$ and $\hat \beta_q^{(\dag)}$ and the corresponding 
mode energies $\lambda_{\pm}(q)$. Therefore the eigenmode occupancies $\hat \alpha_q^{\dag} \hat \alpha_q$ and 
$\hat \beta_q^{\dag} \hat \beta_q$ constitute a set of constants of motion which has as many 
elements as degrees of freedom in our model. 
The occupancies thus form a natural set of operators $\hat I_j$ which 
can be used to set up the density matrix of a GGE Eq.\ (\ref{GGEdensitymatrix}). We again emphasize that in 
Ref.\ \onlinecite{Kollar} an example is given, which shows that an alternative choice of conserved $I_j$ might lead 
to GGE expectation values of observables which differ from the ones obtained by the natural choice. We show 
here that for our model all the studied observables which become stationary in the long time limit approach 
a value consistent with the GGE set up by the set of 
occupancies (called natural GGE below). 

The GGE is described by the density matrix 
\begin{eqnarray}
\label{GGEkonkret}
\hat \rho_{\rm GGE} = \frac{1}{Z_{\rm GGE}} \, e^{- \sum_q \eta_q \hat \alpha_q^\dag \hat \alpha_q -  \sum_q 
\xi_q \hat \beta_q^\dag \hat \beta_q} \; ,
\end{eqnarray} 
with the Lagrange multipliers $\eta_q$ and $\xi_q$ determined by the initial condition
\begin{eqnarray}
\label{ICkonkret}
\left< \hat \alpha_q^\dag \hat \alpha_q \right>_{\hat \rho_{\rm i}} = 
\left< \hat \alpha_q^\dag \hat \alpha_q \right>_{\hat \rho_{\rm GGE}}
\end{eqnarray} 
and similarly for $\hat \beta_q$. For a density matrix of the form Eq. (\ref{GGEkonkret}) the 
eigenmode occupancies are given by\cite{Mermin} 
\begin{eqnarray}
\label{GGeexpec}
\left< \hat \alpha_q^\dag \hat \alpha_q \right>_{\hat \rho_{\rm GGE}} = n_{\rm B}(\eta_q) \; , \;\;
 \left< \hat \beta_q^\dag \hat \beta_q \right>_{\hat \rho_{\rm GGE}} = n_{\rm B}(\xi_q) 
\end{eqnarray}
with the Bose function
\begin{eqnarray*}
n_{\rm B}(x) = \left[  e^x - 1 \right]^{-1} \; .
\end{eqnarray*}

To fix the Lagrange multipliers we still have to compute the left hand side of Eq.\ (\ref{ICkonkret}) for 
the two initial density matrices Eqs.\ (\ref{iniquench}) and (\ref{inik_0}). For the quench we 
obtain using (the inversion of) Eq.\ (\ref{trafo}) 
\begin{eqnarray}
\left< \hat \alpha_q^\dag \hat \alpha_q \right>_{\hat \rho^{\rm q}_{\rm i}} & = & c_q^2 n_{\rm B}(\beta v_{\rm F} q) 
+ s_q^2 n_{\rm B}(\beta \omega_0) \; , \nonumber \\
\label{ICquench} 
\left< \hat \beta_q^\dag \hat \beta_q \right>_{\hat \rho^{\rm q}_{\rm i}} & = & s_q^2 n_{\rm B}(\beta v_{\rm F} q) 
+ c_q^2 n_{\rm B}(\beta \omega_0) \; ,
\end{eqnarray} 
which leads to the set of nonlinear equations for the $\eta_q$ and $\xi_q$
\begin{eqnarray}
n_{\rm B}(\eta_q) & = & c_q^2 n_{\rm B}(\beta v_{\rm F} q) 
+ s_q^2 n_{\rm B}(\beta \omega_0) \; , \nonumber \\
\label{LPfix}
 n_{\rm B}(\xi_q)  & = & s_q^2 n_{\rm B}(\beta v_{\rm F} q) 
+ c_q^2 n_{\rm B}(\beta \omega_0)  \; .
\end{eqnarray}
Obviously, the $\eta_q$ and $\xi_q$ are functions of the inverse temperature $\beta$. 

For the $k_0$-excitation it follows using (the inversion of) Eq.\ (\ref{trafo}) 
and the ``vacuum''-properties of the initial pure state that  
\begin{eqnarray}
\left< \hat \alpha_q^\dag \hat \alpha_q \right>_{\hat \rho^{k_0}_{\rm i}} & = & \mbox{Tr}\, \left[ \hat \rho_{\rm i}^{k_0} 
\hat \alpha_q^\dag \hat \alpha_q  \right] \nonumber \\
& = &  \left<\mbox{vac} \right| \otimes  \left< \mbox{FS} \, \right| \hat a_{k_0} 
\left( c_q \hat b_q^\dag + s_q B_q^\dag \right) \nonumber \\ && \times \left( c_q \hat b_q + s_q B_q \right) 
\hat a_{k_0}^\dag \left| \mbox{FS} \, \right> \otimes \left| \mbox{vac} \right> \nonumber \\ 
& = & c_q^2 \,  \left< \mbox{FS} \, \right|   \hat a_{k_0}  \hat b_q^\dag  \hat b_q \hat a_{k_0}^\dag 
\left| \mbox{FS} \, \right> \nonumber \\
& = &  c_q^2 \,  \left< \mbox{FS} \, \right|   \hat a_{k_0}  \hat b_q^\dag  
\left( \left[\hat b_q , \hat a_{k_0}^\dag  \right] +  \hat a_{k_0}^\dag \hat b_q \right) 
\left| \mbox{FS} \, \right>\nonumber \\
& = &  c_q^2 \,{\frac{2 \pi}{Lq}} \,  
\left< \mbox{FS} \, \right|   \hat a_{k_0-q}  \hat a_{k_0-q}^\dag   \left| \mbox{FS} \, \right> \nonumber \\
\label{ICk_01}
& = & c_q^2  \,{\frac{2 \pi}{Lq}} \, \Theta(k_0-q)  \; ,
\end{eqnarray} 
with the $\Theta$-function defined such that $\Theta(0)=1$. In the step from the fourth to the fifth 
equation we have used twice that from Eqs.\ (\ref{density}) and (\ref{bosons}) it follows 
that 
\begin{eqnarray*}
 \left[\hat b_q , \hat a_{k_0}^\dag  \right] =  \sqrt{\frac{2 \pi}{Lq}} \,  \hat a_{k_0-q}^\dag \; . 
\end{eqnarray*}
Similarly we obtain
\begin{eqnarray}
\label{ICk_02}
\left< \hat \beta_q^\dag \hat \beta_q \right>_{\hat \rho^{k_0}_{\rm i}}  
 =  s_q^2  \,{\frac{2 \pi}{Lq}} \, \Theta(k_0-q)  \; ,
\end{eqnarray} 
which determines the Lagrange parameters using Eqs.\ (\ref{ICkonkret}) and (\ref{GGeexpec}).

\subsection{Expectation values for the quench}

We are now in a position to determine the GGE expectation values for the subsystem energies, the 
phonon momentum distribution $N_{\rm GGE}(q)$, and the fermion momentum distribution 
$n_{\rm GGE}(k)$ first focusing on the quench. With Eqs.\ (\ref{hambos}), (\ref{trafo}), and 
(\ref{ICquench}) we straightforwardly 
obtain 
\begin{eqnarray}
\left< \hat H_{\rm e} \right>_{\hat \rho_{\rm GGE}} & = & \sum_{q>0} v_{\rm F} q 
\left\{  n_{\rm B}(\beta v_{\rm F} q) - 
2  c_q^2  \, s_q^2  \right. \nonumber \\ && \left. \times 
\left[  n_{\rm B}(\beta v_{\rm F} q ) - n_{\rm B}(\beta \omega_0)  
\right]  \right\} \; , \nonumber \\
\left< \hat H_{\rm p} \right>_{\hat \rho_{\rm GGE}} & = & \omega_0 \sum_{q>0} 
\left\{  n_{\rm B}(\beta \omega_0) +
2  c_q^2  \, s_q^2  \right.  \nonumber \\  \nonumber  
&& \left. \times 
\left[  n_{\rm B}(\beta v_{\rm F} q ) - n_{\rm B}(\beta \omega_0)  
\right] 
\right\} \; ,  \\
\left< \hat H_{\rm ep} \right>_{\hat \rho_{\rm GGE}} & = & 2 g  \sum_{q>0} \sqrt{q}  \, 
c_q s_q \left(c_q^2 - s_q^2   \right) \nonumber \\ && \label{GGEenergiesquench} 
\times  \left[  n_{\rm B}(\beta v_{\rm F} q ) - n_{\rm B}(\beta \omega_0)  
\right] \; .
\end{eqnarray} 
The first terms in the subsystem energies are the expectation values taken with 
the initial canonical ensemble at $g=0$. Without 
a high momentum cutoff the energy in the phonon subsystem obviously diverges.
Here we avoid to introduce such a cutoff by considering the excess energies 
$\delta \left< \hat H_{\rm p} \right>_{\hat \rho_{\rm GGE}} =  
\left< \hat H_{\rm p} \right>_{\hat \rho_{\rm GGE}} - 
\left< \hat H_{\rm p} \right>_{\hat \rho^{\rm q}_{\rm i}}$ and $\delta 
\left< \hat H_{\rm e} \right>_{\hat \rho_{\rm GGE}}$ (defined similarly)  in the following. 
Note that the momentum sums containing a factor $s_q$ are cut off at $q_c$ as 
$s_q=0$ for $q>q_c$. With 
\begin{eqnarray*}
\sum_{q > 0} \ldots \to \frac{L}{2\pi} \int_0^\infty dq \ldots   
\end{eqnarray*}
for large $L$ it becomes apparent that the 
subsystem excess energies and the energy in the electron-phonon 
coupling are {\it extensive} and scale $\sim L$. 

For the phonon momentum distribution function it follows similarly that 
\begin{eqnarray}
\delta  N_{\rm GGE}(q) & = & N_{\rm GGE}(q) -   n_{\rm B}(\beta \omega_0) \nonumber
 \\ \label{GGEphoquench} 
& = & 2  c_q^2  \, s_q^2  
\left[  n_{\rm B}(\beta v_{\rm F} q ) - n_{\rm B}(\beta \omega_0)  \right]  \; .
\end{eqnarray}
It is instructive to compare this result to the canonical (equilibrium) phonon 
distribution of the {\it interacting} system (same parameters $g$ and $\omega_0$) 
at some temperature $\tilde T=1/\tilde \beta$ 
characterized by the density matrix Eq.\ (\ref{canny}). A straightforward 
calculation using the methods introduced above gives 
\begin{eqnarray*}
N_{\rm can}(q)  =  s_q^2  n_{\rm B}(\tilde \beta \lambda_+(q)) 
+ c_q^2 n_{\rm B}(\tilde \beta \lambda_-(q))  \; .
\end{eqnarray*}
Obviously, the $q$-dependences of this function and Eq.\ (\ref{GGEphoquench})  
differ and no temperature $\tilde T$ can be found leading to coinciding results.  

\begin{figure}[t]
\includegraphics[width=.8\linewidth,clip]{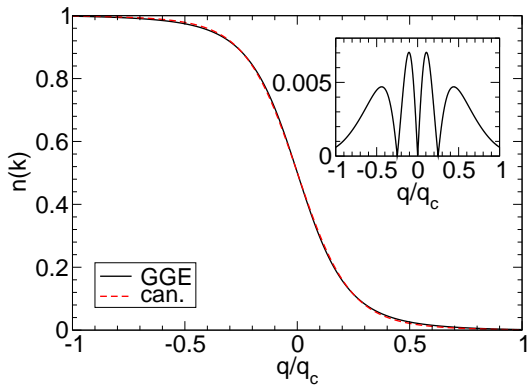}
\caption{(Color online) The GGE and canonical momentum distribution function of the 
fermions. The system parameters are 
$\Gamma=0.01$, $\Omega=0.1$, $\nu=2\pi/(L q_{\rm c})=10^{-3}$ and the dimensionless 
initial temperature 
is $\tau=T/(v_{\rm F} q_{\rm c})=0.1$. The canonical distribution is the best fit 
to the GGE one with 
the fitted temperature $\tilde \tau_{\rm b} =0.10275$. The inset shows the absolute value of the 
difference of the two functions.}
\label{fig2}
\end{figure}

Using the method introduced in Sect. \ref{sec:method}, the Baker-Hausdorff relation, 
and the formula\cite{Mermin} 
\begin{eqnarray*}
\left< e^{\hat A} e^{\hat B} \right> = e^{\left< \hat A^2 + 2 \hat A \hat B + \hat B^2 \right>/2}
\end{eqnarray*}
one obtains for the fermionic momentum distribution function in the GGE
 \begin{eqnarray}
&& \hspace{-.4cm} n_{\rm GGE}(k) =  \frac{1}{L} \int_{-L/2}^{L/2} dx \, e^{-ikx} \,  
\exp{\left\{ \sum_{q>0} \frac{2\pi}{Lq} \, e^{-iq(x-i0)} \right\}} \nonumber \\ 
&& \hspace{-.4cm}  \times \exp{ \left\{ -  \sum_{q>0} \frac{4\pi}{Lq} \left[ 1- \cos(qx) \right] 
n_{\rm B} (\beta v_{\rm F} q) \right\}}  \label{GGEelecquench} \\ 
&&\hspace{-.4cm}  
 \times \exp{ \left\{   \sum_{q>0}  \frac{8\pi}{Lq} \,  c_q^2 s_q^2  \left[ 1- \cos(qx) \right]  \left[   
n_{\rm B}(\beta v_{\rm F} q ) - n_{\rm B}(\beta \omega_0)   \right]    \right\}}  \nonumber  \; .
\end{eqnarray} 
The third factor, which vanishes for $g=0$, contains the information about 
the electron-phonon coupling. 
The first two terms constitute the canonical momentum distribution at temperature $T=1/\beta$ 
for a {\it noninteracting} fermionic system with linear dispersion on a ring of size $L$ in 
equilibrium. This $g=0$ expression 
was earlier derived in Ref.\ \onlinecite{VM5}. As discussed there it becomes equal to the 
(grand canonical) Fermi function $\left[ e^{\beta v_{\rm F} q} +1 \right]^{-1}$ (only) in the 
thermodynamic limit $L \to \infty$. 
Instead of numerically performing the sums and the integral for finite $L$ Eq.\ (\ref{GGEelecquench}) can very efficiently 
be evaluated using an iterative approach introduced in the Appendix of Ref.\ \onlinecite{VM5}. 
Adopting this method to the present situation we obtain for 
$m \in {\mathbb Z}$, $k_m = 2 \pi m/L$, $l=0,1,2,\ldots$   
\begin{eqnarray}
\label{it1}
 && n_{\rm GGE}(k_m)  =  \sum_{n=0}^\infty a_{n+m} \; , \\
&&  a_{l}   =  \exp{ \left\{ -2 \sum_{n=0}^\infty \frac{f(n)}{n} \right\} } \, \sum_{m=0}^\infty c_m c_{m+l} =  a_{-l} \; ,  \nonumber \\
&& c_m  =  \frac{1}{m} \sum_{l=1}^m f(l) c_{m-l} \; ,  \nonumber \\
&& f(l)  =  n_{\rm B} (\beta v_{\rm F} q_l) - 2 c_{q_l}^2 s_{q_l}^2 \left[   
n_{\rm B}(\beta v_{\rm F} q_l ) - n_{\rm B}(\beta \omega_0)   \right] \; , \nonumber
\end{eqnarray}
with $q_l= 2 \pi l/L$. 
As for the phonons we compare this result to the thermal distribution 
with temperature $\tilde T$. The canonical fermion 
distribution function of the interacting system can be computed along the same lines as the GGE 
distribution. Because of the involved structure a comparison of the analytical expressions
is less instructive as for the phonons. In Fig.\ \ref{fig2} we therefore compare numerical 
results for the GGE with $\Gamma=0.01$, $\Omega=0.1$, system size parameter 
$\nu=2\pi/(L q_{\rm c})=10^{-3}$, and dimensionless temperature $\tau=T/(v_{\rm F} q_{\rm c})=0.1$ with 
the best fit of a canonical distribution function for the same parameters---in particular the {\it same} 
electron-phonon coupling and the {\it same} system size---and fitting parameter 
$\tilde \tau= \tilde T/(v_{\rm F} q_{\rm c})$.  The best agreement 
is achieved for $\tilde \tau_{\rm b} =0.10275$. In general $\tilde \tau_{\rm b}$ depends on the model 
parameters and $\tau$. The differences are small but significant as becomes explicit in the inset which 
shows the absolute value of the difference of the two distributions. Doubling the system size 
(halving $\nu$) does not lead to any changes on the scale of the main plot as well as 
the one of the inset. Thus the curves can be considered to be in the thermodynamic limit 
and the two ensembles lead to different results even after the latter has been performed. 
This is a crucial observation as we have to distinguish this type of deviation between 
the predictions of two ensembles from the one which might appear at finite $L$ but vanishes 
for $L \to \infty$. An example for the latter case is the difference between the canonical and 
the grand canonical ensembles as referred to in lectures on statistical mechanics. For 
noninteracting fermions with a linear dispersion such {\it finite size} differences are 
explicitly studied in  Ref.\ \onlinecite{VM5}.   

\subsection{Expectation values for the $k_0$-excitation}

We next derive the same expectation values but for the $k_0$-excitation. 
Using Eqs.\ (\ref{hambos}), (\ref{trafo}), 
(\ref{ICk_01}), and (\ref{ICk_02}) we obtain 
\begin{eqnarray}
\left< \hat H_{\rm e} \right>_{\hat \rho_{\rm GGE}} & = & \sum_{q>0} v_{\rm F} q \, \frac{2 \pi}{L q} \, 
\left( c_q^4 + s_q^4 \right)  \, \Theta(k_0-q) \; ,\nonumber \\
\label{GGEenergiesk_0}
\left< \hat H_{\rm p} \right>_{\hat \rho_{\rm GGE}} & = & 2 \omega_0 \sum_{q>0} \frac{2 \pi}{L q} \, c_q^2 \, 
s_q^2   \, \Theta(k_0-q)  \; , \\
\left< \hat H_{\rm ep} \right>_{\hat \rho_{\rm GGE}} & = & 2 g  \sum_{q>0} \frac{2 \pi}{L q} \, \sqrt{q}  \, 
c_q s_q \left(c_q^2 - s_q^2   \right) \, \Theta(k_0-q) \; . \nonumber 
\end{eqnarray}
In contrast to the quench case the energies are {\it not} extensive. This is related to the fact that even 
for $L \to \infty$ we only add a {\it single} additional fermion at momentum $k_0$ to the filled Fermi 
sea.   
The phonon momentum distribution function is given by
\begin{eqnarray}
\label{GGEphok_0}
N_{\rm GGE}(q) = 2  \frac{2 \pi}{L q} \, c_q^2 \, 
s_q^2   \, \Theta(k_0-q)  \; .
\end{eqnarray}
Using the bosonization of the fermionic fields the GGE expectation value for the 
momentum distribution function follows as 
 \begin{eqnarray*}
&& \hspace{-.4cm} n_{\rm GGE}(k) =  \frac{1}{L} \int_{-L/2}^{L/2} dx \, e^{-ikx} \,  
\exp{\left\{ \sum_{q>0} \frac{2\pi}{Lq} \, e^{-iq(x-i0)} \right\}} \nonumber \\ 
&& \hspace{-.4cm} \times \exp{ \left\{ 2  \sum_{q>0} \left(\frac{2\pi}{Lq}\right)^2
\left( c_q^4 + s_q^4 \right) \left[ 1- \cos(qx) \right] \Theta(k_0-q) \right\} }  \; .
\end{eqnarray*} 
It is now crucial to realize that due to the factor $1/L^2$ in the exponent the 
second term approaches 
1 in the thermodynamic limit $L \to \infty$ independent of the electron-phonon coupling. 
In this limit the remaining terms form a step function. We thus find
 \begin{eqnarray}
\lim_{L \to \infty} n_{\rm GGE}(k) = \Theta(k)    
\end{eqnarray}  
and the fermionic momentum distribution function of the GGE in the thermodynamic 
limit becomes equal to the one of the {\it groundstate} (which for the present model 
is equal to the noninteracting one; see above). This is consistent with
the observation that the energies for the $k_0$-excitation are not 
extensive. For finite $L$, $ n_{\rm GGE}(k) $ can again be computed
iteratively using Eq.\ (\ref{it1}) with $f(l)$ replaced by 
\begin{eqnarray}
 \label{it2}
f(l) =  \left( c_{q_l}^4 + s_{q_l}^4 \right) \Theta(k_0-q_l)/l \; . 
\end{eqnarray} 
We note in passing that as for the quench for finite $L$, $n_{\rm GGE}(k) $  
is {\it different} from the canonical distribution function obtained for
 the same system parameters at an optimally chosen temperature 
$\tilde T$. The same holds for $N_{\rm GGE}(q)$.

\section{The quench dynamics}
\label{sec:quench}

We now investigate the dynamics of the density matrix and the expectation values considered 
above under the Hamiltonian Eq.\ (\ref{hamiltonian1}) starting out from the 
{\it noninteracting} ($g=0$) canonical density matrix Eq.\ (\ref{iniquench}) 
at temperature $T=1/\beta$. Using the formal solution
\begin{eqnarray}
\label{solvonneu}
\hat \rho^{\rm q}(t) = e^{-i \hat H t} \, \hat \rho^{\rm q}_{\rm i} \, e^{i \hat H t}
\end{eqnarray}
of the von Neumann equation and Eq.\ (\ref{trafo}) 
we end up with 
\begin{eqnarray}
Z \, \hat \rho^{\rm q}(t)  =  \exp && \!\!\!\!\! \left\{     
- \beta \sum_{q>0} \left[ \phantom{ \hat \beta_q^\dag} \!\!\!\!\!\!  
\left( v_{\rm F} q c_q^2 + \omega_0 s_q^2 \right) \hat \alpha_q^\dag \hat \alpha_q  \right. \right. \nonumber \\
&& + c_q s_q \left(\omega_0 -v_{\rm F} q \right) e^{- i \Delta \lambda(q) t } 
\hat  \alpha_q^\dag \hat \beta_q  + \mbox{H.c.}   
\nonumber \\
\label{rhoqt}
&& \left. \left.  + \left( v_{\rm F} q s_q^2 + \omega_0 c_q^2 \right) \hat \beta_q^\dag \beta_q  
\right] \phantom{\sum_{q>0}} \!\!\!\!\!\!\!\!\!\!  \right\} \; ,
\end{eqnarray}
where
\begin{eqnarray*}
\Delta \lambda(q)  = \lambda_+(q) - \lambda_-(q) \; .
\end{eqnarray*}
We emphasize that only the difference of the two eigenmode energies enter the dynamics. To compute the 
expectation values of interest with $\hat \rho^{\rm q}(t)$ one can now diagonalize the {\it time dependent} 
$2\times 2$ coefficient matrix of the quadratic form appearing in the 
exponent of Eq.\ (\ref{rhoqt}) 
and introduce new {\it time dependent} bosonic operators as linear combinations
of the $\hat \alpha_q^{(\dag)}$ and $\hat \beta_q^{(\dag)}$. Alternatively, one can use Eqs.\ (\ref{Bqt}) and
(\ref{bqt}) and compute the expectation values with  $\hat \rho^{\rm q}_{\rm i}$. 

For the time dependence of the subsystem excess energies and interaction energy this leads to 
\begin{eqnarray}
\delta \left< \hat H_{\rm e} \right>_{\hat \rho^{\rm q}(t)}  & = & - \sum_{q>0} v_{\rm F} q 
2  c_q^2  \, s_q^2  \nonumber \\ 
&& \hspace{-1.7cm}  \times \left[  n_{\rm B}(\beta v_{\rm F} q ) - n_{\rm B}(\beta \omega_0) \right] \left[ 
1- \cos \left\{ \Delta \lambda(q) t \right\} \right] \; , \nonumber \\
\delta \left< \hat H_{\rm p} \right>_{\hat \rho^{\rm q}(t)}  & = &  
\omega_0 \sum_{q>0}  2  c_q^2  \, s_q^2  \nonumber \\ 
&& \hspace{-1.7cm}  \times \left[  n_{\rm B}(\beta v_{\rm F} q ) - n_{\rm B}(\beta \omega_0)  \right]  \left[ 
1- \cos \left\{ \Delta \lambda(q) t \right\} \right] \; , \nonumber \\
\left< \hat H_{\rm ep} \right>_{\hat \rho^{\rm q}(t)} 
& = & 2 g  \sum_{q>0}  \sqrt{q} \, c_q s_q \left(c_q^2-s_q^2\right)  \nonumber \\  \label{energiesquench}
&&  \hspace{-1.7cm} \times \left[  n_{\rm B}(\beta v_{\rm F} q ) - n_{\rm B}(\beta \omega_0)  \right]  \left[ 
1- \cos \left\{ \Delta \lambda(q) t \right\} \right]  \; .
\end{eqnarray}
It is easy to show that for all $t$ the total excess energy $ \delta \left< \hat H 
\right>_{\hat \rho^{\rm q}(t)}$ sums up to zero due to energy conservation. 
To answer the question if the (excess) energies become stationary in the long-time limit we {\it first}
perform the  {\it thermodynamic limit.} Afterwards the oscillatory terms 
average out for $t  \to \infty$ when the momentum {\it integrals} are performed. 
For $L \to \infty$ the (excess) energy expectation values (per particle) thus become 
{\it stationary} and {\it equal} to the GGE expectation values (per particle) as determined 
in Eq.\ (\ref{GGEenergiesquench}). This provides a first indication that for our model 
long-time expectation values of observables which become stationary can indeed be computed using the 
appropriate GGE. 
As discussed in the introduction in performing the thermodynamic limit one often keeps 
the size of one of the subsystems (the $\mathcal S$) containing the relevant local observables 
fixed while the thermodynamic limit is performed in its complement, the 
environment $\mathcal E$.\cite{Barthel}  
In our model the limit $L \to \infty$ is simultaneously performed in the fermion and 
phonon subsystems (quantization of momenta). 

\begin{figure}[tb]
\includegraphics[width=.9\linewidth,clip]{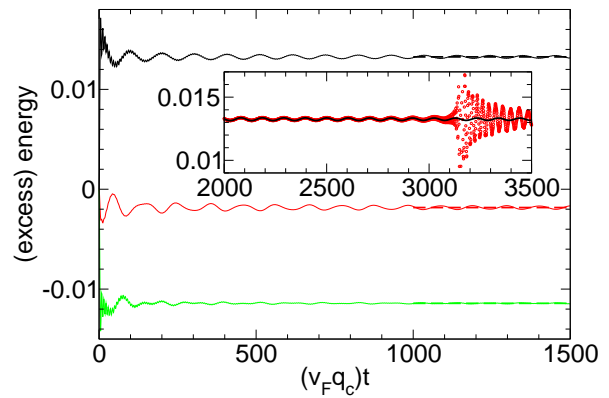}
\caption{(Color online) Time evolution of the (excess) energies in the fermion (solid line, top) and phonon 
(solid line, middle) subsystems 
as well as of the energy in the interacting part of the Hamiltonian (solid line, bottom). The dashed lines 
for $t \in [1000,1500]$ are the corresponding GGE expectation values. 
The parameters are $\Gamma=0.01$, $\Omega=0.1$ and the 
initial temperature is given by $\tau=0.1$. The symbols in the inset show data for the energy in the 
fermion subsystem obtained at a finite system size $\nu=2 \times 10^{-3}$. The recurrence time is 
then given by $(v_{\rm F} q_{\rm c}) t_r = v_{\rm F} 2 \pi /(L \nu) L/v_{\rm F} = 2 \pi / \nu 
\approx 3142$.}
\label{fig3}
\end{figure}

As shown in Fig.\ \ref{fig3} (solid lines) for large $t$ the energies oscillate 
around their asymptotic GGE values (dashed lines) with seemingly a single frequency and a slowly decaying 
amplitude. From the data it is obvious that a rather good approximation for the stationary expectation 
values can be obtained by averaging the (excess) energies over a time interval which is much larger 
than the inverse oscillation frequency and starts at a sufficiently large time.    

Analytic insights on the long-time behavior can be gained by applying the techniques of  
{\it asymptotic analysis}\cite{Bender} such as the {\it stationary phase method} to the momentum integrals in 
Eq.\ (\ref{energiesquench}). Using the latter 
one can show that for not too strong couplings $2\Gamma< \Omega$ (on which we mainly focus) such 
that the stationary point (of $\Delta \lambda(q)$) lies inside the integration interval $(0,q_c)$ 
the dominant oscillation frequency (at large times) is given by the {\it minimum} of the mode 
energy difference 
\begin{eqnarray*}
\frac{\Delta \lambda(q_{\rm min})}{v_{\rm F} q_{\rm c}} = 2 \Gamma \sqrt{\frac{\Omega}{\Gamma} -1}
\end{eqnarray*}
and the amplitude falls off as $1/\sqrt{t}$.\cite{Fioretto} The convergence towards the 
stationary values is thus rather slow.   
Furthermore, the second (high) frequency visible at small times 
results from a contribution of the boundaries of the momentum integrals and can 
be identified as $\Delta \lambda(q_c)$. The mode energy difference of the lower boundary 
does not appear as a frequency as the corresponding amplitude vanishes.

For comparison the inset of Fig.\ \ref{fig3} also contains results for the energy in the electron subsystem 
obtained at {\it finite} $L$ (symbols). Obviously on the scale $t_{\rm r} = L/v_{\rm F}$ finite size 
effects set in and the oscillations become rather erratic. Remarkably, for upper times 
not too large compared to $t_{\rm r}$ the $L \to \infty$ asymptotic value can still 
be extracted accurately by averaging over the time even beyond $t_{\rm r}$. The same holds for the 
other energy expectation values. 

\begin{figure}[tb]
\includegraphics[width=.95\linewidth,clip]{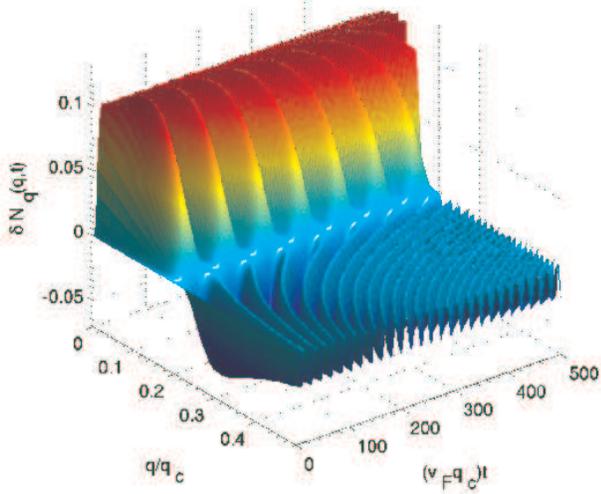}
\caption{(Color online)  The time-dependent part of the phonon momentum distribution function 
$\delta  N_{\rm q}(q,t)$ as a function of $q$ and $t$. The parameters are $\Gamma=0.01$, $\Omega=0.2$, 
and the initial temperature is given by $\tau=0.1$.}
\label{fig4}
\end{figure}

From the energy in the phonon subsystem Eq.\ (\ref{energiesquench}) the time evolution of  
the phonon momentum distribution function out of the initial Bose function $n_{\rm B}(\beta \omega_0)$ 
can be read off
\begin{eqnarray}
\delta  N_{\rm q}(q,t) & = & N_{\rm q}(q,t)-n_{\rm B}(\beta \omega_0) \nonumber \\
& = & 2 c_q^2  \, s_q^2  \left[  n_{\rm B}(\beta v_{\rm F} q )
 - n_{\rm B}(\beta \omega_0)  \right]  \nonumber \\ 
\label{quenchtimeevolbos}
&& \times \left[ 
1- \cos \left\{ \Delta \lambda(q) t \right\} \right] \; .
\end{eqnarray}
As the phononic annihilation and creation operators are {\it linear} combinations of the 
eigenmode ladder operators (see Eq.\ (\ref{trafo})) it is not surprising, that the phonon 
momentum 
distribution function for a fixed $q$ does not become stationary; it shows a sinusoidal 
oscillation with frequency $\Delta  \lambda(q)$ and fixed amplitude for all $t$. 
The time-dependent part $\delta  N_{\rm q}(q,t) $ is shown in Fig.\ \ref{fig4} 
for $\Gamma=0.01$, $\Omega=0.2$, and temperature $\tau = 0.1$. The shape for fixed $t$ can 
be understood from Fig.\ \ref{fig1} (in which $s_q^2$ and $c_q^2$ are shown) and the $q$-dependence 
of the difference of the two Bose functions. 
Only after averaging over a small interval $\Delta q$ around $q$, $N_{\rm q}(q,t)$ 
becomes stationary at large 
times as the cosine term drops out. 

This behavior has to be contrasted to the one of the fermion momentum distribution 
function $n_{\rm q}(k,t)$. By the bosonization of the field operator (see Sect. \ref{sec:method}) 
the relation between the $\hat a_k^{(\dag)}$ and the eigenmode ladder operators is highly 
nonlinear and it is not clear a priori whether or not $n_{\rm q}(k,t)$ becomes stationary.
Using the methods introduced above we obtain 
\begin{eqnarray}
&& \hspace{-.4cm} n_{\rm q}(k,t) =  \frac{1}{L} \int_{-L/2}^{L/2} dx \, e^{-ikx} \,  
\exp{\left\{ \sum_{q>0} \frac{2\pi}{Lq} \, e^{-iq(x-i0)} \right\}} \nonumber \\ 
&&  \hspace{-.4cm} \times \exp{ \left\{ - \sum_{q>0} \frac{4\pi}{Lq} \left[ 1- \cos(qx) \right] 
n_{\rm B} (\beta v_{\rm F} q) \right\}}  \label{quenchtimevolferm} \\ 
&&  \hspace{-.4cm} \times \exp 
\left\{ 
 \sum_{q>0}  \frac{8\pi}{Lq} \,  c_q^2 s_q^2  \left[ 1- \cos(qx) \right]  \left[   
n_{\rm B}(\beta v_{\rm F} q ) - n_{\rm B}(\beta \omega_0)   \right]   \nonumber \right. \\
&&  \hspace{-.4cm} \left. \;\;\;\;\;  \;\;\;\;\;  \phantom{\sum_{q>0}   \frac{2\pi}{Lq} } \times 
\left[ 
1- \cos \left\{ \Delta \lambda(q) t \right\} \right] 
 \right\}  \nonumber  \; .
\end{eqnarray} 
First taking the thermodynamic limit and subsequently the long-time
limit the oscillatory time dependent term again averages out and we find {\it convergence} to the
GGE expectation value Eq.\ (\ref{GGEelecquench}), 
\begin{eqnarray*}
\lim_{t \to \infty} \; \lim_{L \to \infty} n_{\rm q}(k,t) =  \lim_{L \to \infty} 
n_{\rm GGE}(k) \; .
\end{eqnarray*}
Furthermore, an analytic  stationary phase analysis similar to the one performed for the 
energies shows that for sufficiently large $t$ and fixed $k$, $n_{\rm q}(k,t)$ oscillates
with frequency $\Delta \lambda(q_{\rm min})$ around $n_{\rm GGE}(k)$ with an amplitude
which decays as $1/\sqrt{t}$. As for the energies the second relevant frequency  
is $\Delta \lambda(q_{\rm c})$.

\begin{figure}[tb]
\includegraphics[width=.8\linewidth,clip]{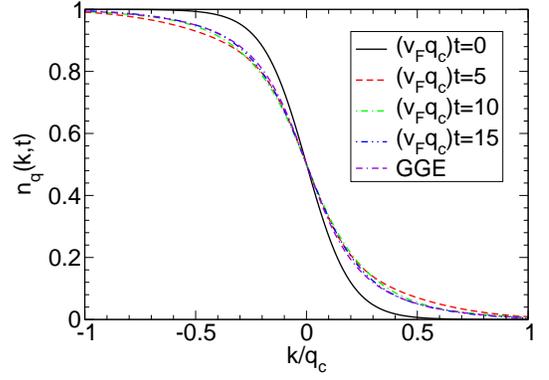}
\caption{(Color online) Momentum distribution function of the fermions as a function of 
$k$ for different $t$. The GGE distribution obtained for the same system size is shown 
for comparison. The parameters are $\Gamma=0.03$, $\Omega=0.1$, $\nu=10^{-3}$,  
and the initial temperature is given by $\tau=0.1$. }
\label{fig5}
\end{figure}

\begin{figure}[tb]
\includegraphics[width=.8\linewidth,clip]{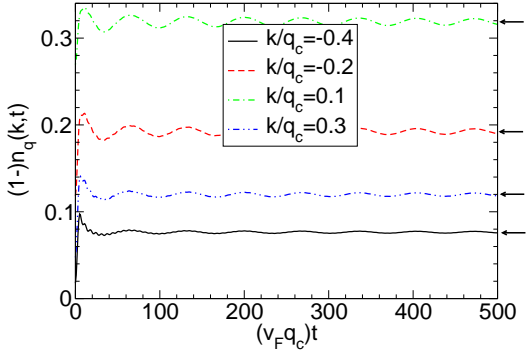}
\caption{(Color online) The same as in Fig.\ \ref{fig5}, but  
as a function of $t$ for different $k$. For the momenta 
$k<0$, $1-n_{\rm q}(k,t)$ is shown. The GGE expectation values 
are indicated by the arrows.}
\label{fig6}
\end{figure}

For finite (but large) $L$, $n_{\rm q}(k,t)$ can again 
efficiently be computed numerically using the recursion relation Eq.\ (\ref{it1}) with $f(l)$ replaced
by 
\begin{eqnarray}
f(l) & = &  n_{\rm B} (\beta v_{\rm F} q_l) - 2 c_{q_l}^2 s_{q_l}^2 \left[   
n_{\rm B}(\beta v_{\rm F} q_l ) - n_{\rm B}(\beta \omega_0)   \right] \nonumber \\
 \label{it3} && \times \left[ 
1- \cos \left\{ \Delta \lambda(q_l) t \right\} \right] \; .
\end{eqnarray} 
In Fig.\ \ref{fig5} the time evolution of the fermion momentum distribution function as a function 
of $k$ and a few $t$ is compared to the GGE prediction obtained for the same system length $L$. 
In Fig.\ \ref{fig6} we show
$n_{\rm q}(k,t)$ for a few fixed $k$ as a function of $t$. 
Almost independent of the system parameters, the temperature, and the considered momentum 
the asymptotic behavior analytically described above (for $L \to \infty$) sets in very 
quickly and holds for times $t<t_{\rm r}$ (oscillation frequencies and $1/\sqrt{t}$ decay).
Just as for the energies the asymptotic values
of $n_{\rm q}(k,t)$ for large $t$ and fixed $k$ can be determined very accurately 
by averaging 
over an appropriate time interval. Similar to the behavior found for the energies the oscillations of 
$n_{\rm q}(k,t)$ for fixed $k$ become rather erratic if one exceeds $t_{\rm r}$ (not shown here), 
but time averaging still gives very good agreement with the GGE result.  
Doubling the system size for the results shown in Figs.\ \ref{fig5} and  \ref{fig6}  does not lead 
to any changes on the scale of the plots and the curves can considered to be in the 
thermodynamic limit (as long as $t< t_{\rm r}$ which is clearly the case in the 
figures; $(v_{\rm F} q_{\rm c})t_{\rm r} \approx 6284$). 

  
We can conclude that for the quench and the observables of interest in the present work, which 
relax to a stationary value, the latter is equal to the GGE prediction. The GGE expectation value differs
from a thermal one (see the discussion in Sect.\ \ref{sec:GGE}, in particular Fig. \ref{fig2}). In the long-time 
limit the expectation values of the 
(subsystem) energies and the momentum distribution function of the fermions at every (fixed) $k$ 
oscillate around the GGE result with a frequency $\Delta \lambda(q_{\rm min})$ and an amplitude 
decaying as $1/\sqrt{t}$. In this respect the dynamics is rather simple---in particular compared 
to the one resulting from the $k_0$-excitation discussed in the next section.  

We briefly comment on 
the relation of our calculations to those performed in the bosonization approach to 
the purely electronic 
Tomonaga-Luttinger model (in and out of equilibrium). For this model  
$q$-sums (-integrals) of the type appearing in Eq.\ (\ref{quenchtimevolferm}) 
are usually performed analytically after {\it Taylor expanding} the renormalized bosonic 
dispersion (here $\Delta \lambda(q)$) to {\it linear order} in $q$ and 
{\it assuming} a particular $q$-dependence of the $s_q^2$ ($=1-c_q^2$). It is generally 
believed that these steps do not alter the low-energy physics of the 
model. However, in Ref.\ \onlinecite{VM6} it was shown that this is not 
correct and the typical Tomonaga-Luttinger liquid exponents are affected by these 
approximations. In particular, this implies that the validity of the 
very interesting study of  the relaxation dynamics of Ref.\ \onlinecite{Cazalilla} 
is more restricted than it is realized by the author(s). For the present model 
it is obvious that similar approximations would strongly alter the physics as 
the nonlinearity of  $\Delta \lambda(q)$ lies at the heart of our results.

\section{The dynamics of the $k_0$-excitation }
\label{sec:k_0_ex}

Using the methods introduced above for the $k_0$-excitation the expectation values of 
the observables studied here have been computed in Ref.\ \onlinecite{VM1}. We emphasize that 
in this paper only the short-time behavior was investigated while we are (mainly) interested
in the long-time asymptotics. For completeness we here present the relevant expressions
taken from Ref.\ \onlinecite{VM1}.

The time-dependence of the (subsystem) energies is given by 
\begin{eqnarray}
\left< \hat H_{\rm e} \right>_{\hat \rho^{k_0}(t)} & = & \sum_{q>0} v_{\rm F} q \, \frac{2 \pi}{L q} \, 
\left[ c_q^4 + s_q^4 + 2 c_q^2 s_q^2 \right. 
\nonumber \\
&& \left. \times \cos \left\{ \Delta \lambda(q) t\right\}
\right]  \, \Theta(k_0-q) \; ,\nonumber \\
\left< \hat H_{\rm p} \right>_{\hat \rho^{k_0}(t)} & = & 2 \omega_0 \sum_{q>0} \frac{2 \pi}{L q} \, c_q^2 \, 
s_q^2   \nonumber \\ 
\label{tevolenergiesk_0}
&& \times  \left[ 1-  \cos \left\{ \Delta \lambda(q) t\right\} \right] \Theta(k_0-q)  \; , \\
\left< \hat H_{\rm ep} \right>_{\hat \rho^{k_0}(t)} & = & 2 g  \sum_{q>0} \frac{2 \pi}{L q} \, \sqrt{q}  \, 
c_q s_q \left(c_q^2 - s_q^2   \right) \nonumber \\
&&  \times  \left[ 1-  \cos \left\{ \Delta \lambda(q) t\right\} \right] \Theta(k_0-q) \; . \nonumber 
\end{eqnarray}
After the thermodynamic limit has been performed in the long-time limit 
the energy expectation values approach the GGE ones Eq.\ (\ref{GGEenergiesk_0}) 
in an oscillatory fashion with the dominant frequency $\Delta \lambda(q_{\rm min})$ and an 
amplitude decaying as $1/\sqrt{t}$. This can be shown analytically 
applying the same methods as used for the quench.
A typical example for the time evolution of the energies is given in Fig.\ 1 of Ref.\ \onlinecite{VM1}. 

For the phonon dynamics one obtains 
\begin{eqnarray}
\label{tevolphok_0}
N_{k_0}(q,t) = 2  \frac{2 \pi}{L q} \, c_q^2 \, 
s_q^2  \left[ 1-  \cos \left\{ \Delta \lambda(q) t\right\} \right]  \Theta(k_0-q)  \; .
\end{eqnarray}
For the reason discussed in the last section the phonon momentum 
distribution function does not become stationary. A plot of $N_{k_0}(q,t)$ similar to our 
Fig.\ \ref{fig4} (obtained for the quench) is shown in Fig.\ 2 of Ref.\ \onlinecite{VM1}. 

To compute $n_{k_0}(k,t)$ for the initial state containing a highly excited fermion in addition to the filled 
Fermi sea one has to bosonize {\it four} fermion fields instead of two. This leads to a rather 
involved expression for the fermion momentum distribution function as discussed in detail in 
the Appendix of  Ref.\ \onlinecite{VM1}. 
In contrast to the quench this expression does not provide much analytical insight and in particular
does not allow to analytically conclude, that 
in the long-time limit $n_{k_0}(k,t)$ becomes equal to the GGE expectation value (at least 
after taking $L \to \infty$ first). We thus have to rely on numerical comparisons.
Therefore here we only give the equations needed for an iterative numerical 
calculation of $n_{k_0}(k,t)$ for finite $L$:\cite{VM1} 
\begin{eqnarray}
\label{edis}
n_{k_0}(k_n,t) & = &  \sum_{r=\mbox{\footnotesize
max}(n,0)}^{n_0}
\sum_{s=0}^{n_0-r} \sum_{l=0}^{\mbox{\footnotesize min}(r+s,r-n)}
a_{r+s-l}^{(n_{\rm c})}(t) \nonumber \\
&& \times \left( a_r^{(n_{\rm c})}(t) \right)^{\ast} b_l^{(n_{\rm c})}(t) \left( b_s^{(n_{\rm c})}
(t) \right)^{\ast}  \; ,
\end{eqnarray}
for $n_0 \geq n > 0$ and $n_{\rm c}=\frac{q_{\rm c} L}{2 \pi}$.
The
time dependent coefficients $a_n^{(m)}$ and $b_n^{(m)}$ are determined  
for $m >1 $, $l \in {\mathbb N}_0$, and $i=0,...,m-1$ by

\parbox{6.0cm}
{\begin{eqnarray*}
a_{lm+i}^{(m)} (t) & = & \sum_{j=0}^{l} \frac{\left[u_m(t)/m
\right]^j}{j!} a_{m(l-j)+i}^{(m-1)}(t) \; , \\
b_{lm+i}^{(m)} (t) & = & \sum_{j=0}^{l} \frac{\left[-u_m(t)/m
\right]^j}{j!} b_{m(l-j)+i}^{(m-1)}(t) \; ,
\end{eqnarray*}}
\hfill
\parbox{2.0cm}
{\begin{equation} \label{coeff} \end{equation}}

\noindent with the starting values ($m=1$) 

\parbox{6.0cm}
{\begin{eqnarray*}
a_{l}^{(1)} (t) & = & \sum_{k=0}^{l} [e(t)]^{l-k} \frac{\left[u_1(t)
\right]^k}{k!} \;\; , \;\; l \in {\mathbb N}_0 \; , \\
b_{l}^{(1)} (t) & = & \frac{\left[-u_1(t)\right]^l}{l!}
- \frac{\left[-u_1(t)\right]^{l-1}}{(l-1)!} e(t) \;\; , \;\;  l \in 
{{\mathbb N}} \; , \\
b_0^{(1)} (t) & = & 1 \; .
\end{eqnarray*}}
\hfill
\parbox{2.0cm}
{\begin{equation} \label{coeffstart} \end{equation}}

\noindent The functions $u_l(t)$ and $e(t)$ are defined as

\parbox{6.0cm}
{\begin{eqnarray*}
u_l(t) & = & c_{q_l}^2 e^{-i \lambda_+(q_l)t}+s_{q_l}^2 e^{-i
\lambda_-(q_l)t}-e^{-i v_{\rm F} q_l t} \; , \\
e(t) & = & e^{-i v_{\rm F} t 2\pi /L} \; .
\end{eqnarray*}}
\hfill
\parbox{2.0cm}
{\begin{equation} \label{fundef} \end{equation}}

\noindent In contrast to all our earlier expressions for the time dependence of expectation values
both eigenenergies $\lambda_{\pm}(q)$ enter explicitely and not only their difference 
$\Delta \lambda(q)$. This leads to very rich dynamics. 

\begin{figure}[t]
\includegraphics[width=.95\linewidth,clip]{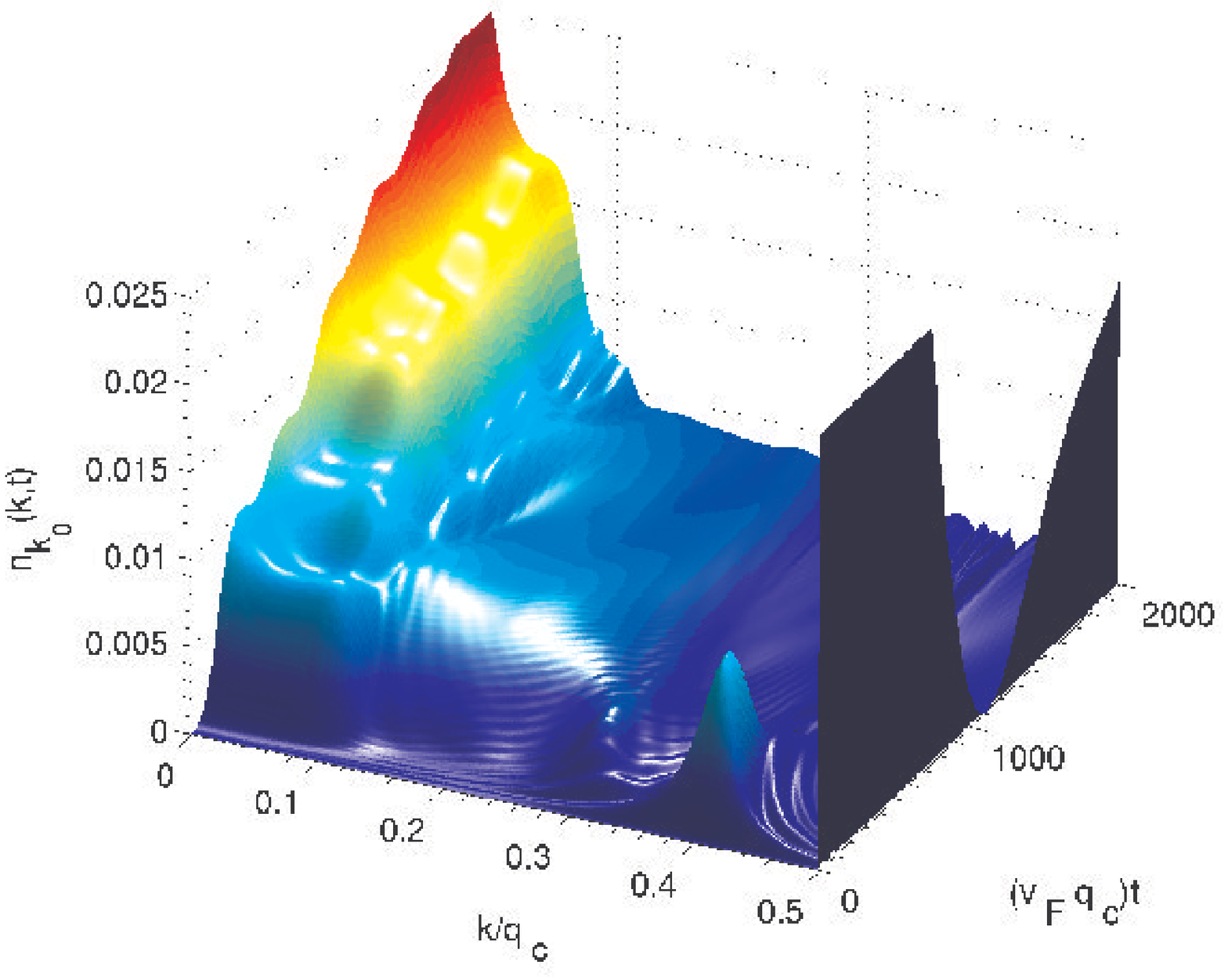}
\caption{(Color online) The momentum distribution function of the fermions 
$n_{k_0}(k,t)$. The parameters are $\Gamma=0.001$, $\Omega=0.1$, $\nu=10^{-3}$, and
$k_0/q_c=0.5$. The dark feature at $k=k_0$ is the occupation of the intially filled level 
which strongly exceeds the scale of the $z$-axis. Note 
the scale of the $z$-axis.}
\label{fig7}
\includegraphics[width=.95\linewidth,clip]{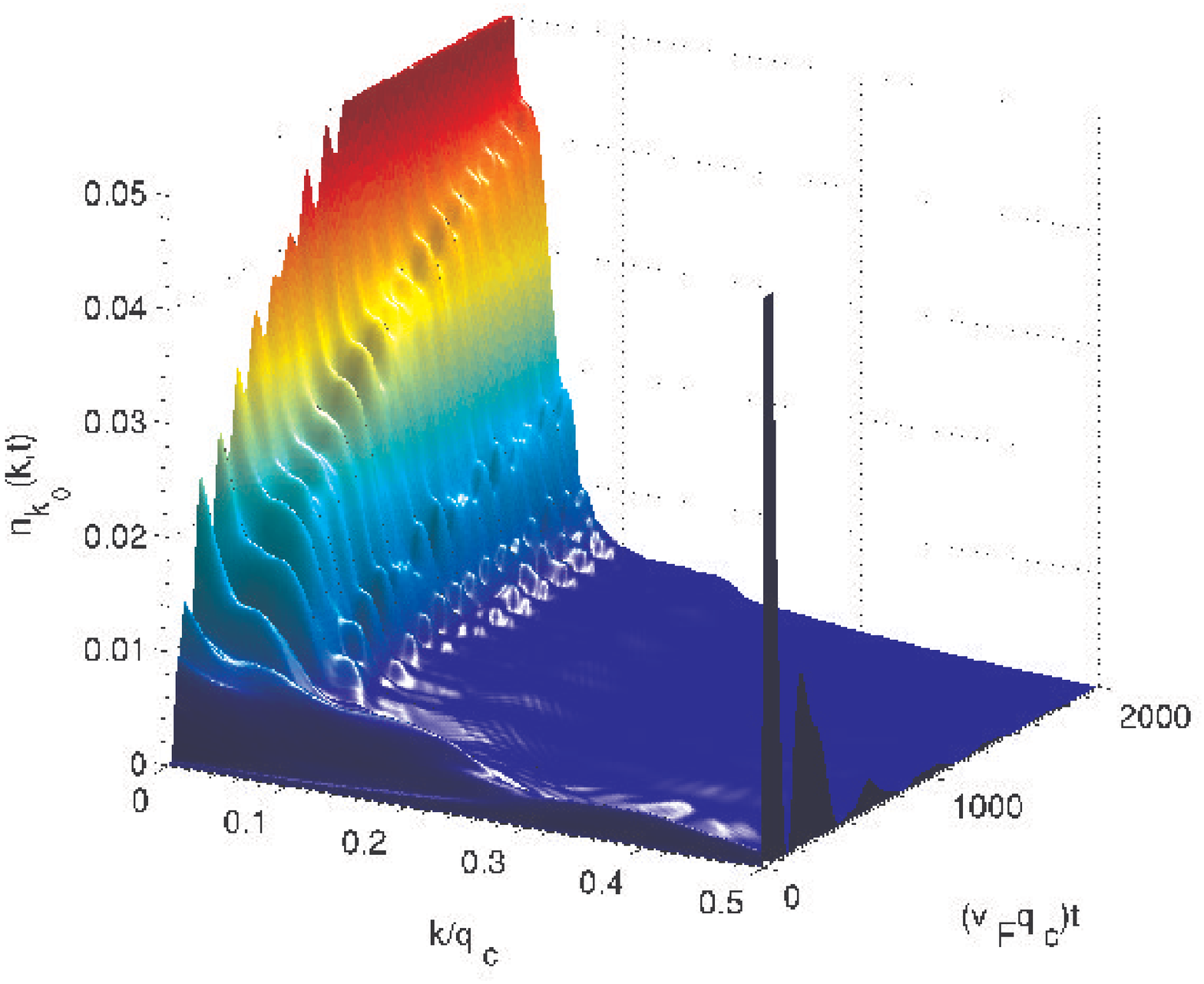}
\caption{(Color online) The same as in Fig.\ \ref{fig7} but for $\Gamma=0.01$.}
\label{fig8}
\end{figure}

Figures \ref{fig7} and \ref{fig8} show two examples of $n_{k_0}(k,t)$ for a weak ($\Gamma=0.001$) and 
an intermediate ($\Gamma=0.01$) electron-phonon coupling at finite but large $L$. Only the part 
$0 \leq k \leq k_0$ is shown; for the time-evolution of initially filled momenta see 
Figs.\ \ref{fig10} and \ref{fig11}. At $t=0$ only the momenta $k<0$ and $k_0$ are occupied (with weight 1) 
and no phonons
are present. The ``hot'' electron decays into states with lower energy by producing phonons. 
The energy conservation in individual scattering processes is not sharp on short time 
scales which leads to {\it broadened replicas} of the initial $k_0$-excitation at momenta 
$k_0-n q_{\rm B}$, with $n \in {\mathbb N}$ and $q_{\rm B}=\omega_0/v_{\rm F}$. 
This is reminiscent to the derivation of Fermi's Golden Rule. In time-dependent 
perturbation theory the transition probability 
between two energy eigenstates with energy difference $\Delta \omega$ is proportional 
to $[\sin(\Delta \omega t /2)/(\Delta \omega/2)]^2$ which only at large $t$ becomes an 
energy conserving $\delta$-function.   
Thus the replicas sharpen for larger times. For our model they never   
reach the width of the original excitation before they get depleted again. Details of this 
dynamics are discussed in Ref.\ \onlinecite{VM1}.  
Two of such replica are visible 
in Fig.\ \ref{fig7} for small $g$, while only a very broad feature appears for intermediate 
couplings (see Fig.\ \ref{fig8}). As soon as phonons are generated they couple to fermions in 
the filled Fermi sea (at $k<0$; not shown in the figures) and excite them to higher energies. 
This leads to a step like feature of width $q_{\rm B}$ at small $k>0$ and $t>0$. 
For increasing  time the sharp initial jump at $k=0$ from 1 to 0 is smoothened. In particular,
the occupancies of the levels at $k \gtrapprox 0$ which are 0 at $t=0$ will increase significantly.
In Figs. \ref{fig7} and \ref{fig8} this is reflected in the continuing overall  increase of 
$n_{k_0}(k,t)$ for small fixed $k$ on time scales ($(v_{\rm F} q_{\rm c})t \approx 10^3$) at 
which saturation is already clearly established for the quench dynamics (see Figs.\ \ref{fig5} 
and \ref{fig6}). 
The initially filled level at $k_0$ is depleted but subsequently refilled on a time scale which 
obviously depends on the electron-phonon coupling (compare Figs.\ \ref{fig7} and \ref{fig8}). For 
large times $n_{k_0}(k_0,t)$ approaches a small value in an oscillatory fashion. 
For the quench the frequency with which $n_{\rm q}(k,t)$ oscillates (at fixed $k$) is 
independent of $k$ and given by $\Delta \lambda(q_{\rm min})$.
In contrast from  Figs.\ \ref{fig7} and \ref{fig8} it is obvious that for the present 
initial nonequilibrium state (and on the same time scales as for the quench) different frequencies
appear (compare the behavior at small $k$ and $k$ close to $k_0$; see also Fig.\ \ref{fig11}). 
We conclude that the dynamics generated by the $k_0$-excitation 
is significantly richer then the one found after a quench.

\begin{figure}[t]
\includegraphics[width=.8\linewidth,clip]{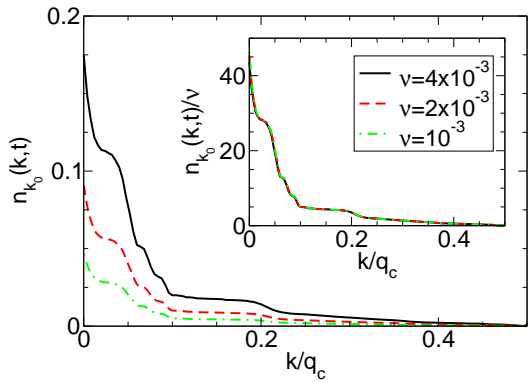}
\caption{(Color online) The momentum distribution function of the fermions 
$n_{k_0}(k,t)$ as a function of $k>0$ for for fixed $(v_{\rm F} q_{\rm c}) t=10^3$ 
and different $\nu=4 \times 10^{-3}$, $2 \times 10^{-3}$, $10^{-3}$. The other parameters 
are $\Gamma=0.01$, $\Omega=0.1$, and $k_0/q_c=0.5$. The inset shows $n_{k_0}(k,t)/\nu$.}
\label{fig9}
\end{figure}

Before comparing numerical data for $n_{k_0}(k,t)$ to the GGE result $n_{\rm GGE}(k)$ we 
have to gain a detailed understanding of the subtleties of the thermodynamic limit for 
the present nonequilibrium initial state.  We already noted that regardless of the 
electron-phonon coupling in the thermodynamic limit 
$n_{\rm GGE}(k)$ is given by  the ground state expectation value; the noninteracting step function. 
This result is consistent with our finding that the excess energies (both the GGE ones and the time-evolved 
ones) are not extensive. We only add the $L$-independent energy $v_{\rm F} k_0$ to the energy of 
the filled Fermi 
sea (the ground state). As the energy is conserved the three terms 
Eq.\ (\ref{tevolenergiesk_0}) add up to $v_{\rm F} k_0$ for all $t \geq 0$. 
Based on these considerations and our previous results we expect that  
\begin{eqnarray*}
\lim_{t \to \infty} \lim_{L \to \infty} n_{k_0}(k,t) = \Theta(k)
\end{eqnarray*}
in accordance with the GGE prediction. 
This is confirmed by Fig.\ \ref{fig9} which shows $n_{k_0}(k,t)$ as a function of $k \geq 0$ 
for fixed $(v_{\rm F} q_{\rm c}) t=10^3$ and different $\nu=4 \times 10^{-3}$, $2 \times 10^{-3}$, 
$10^{-3}$. In the inset we plot $n_{k_0}(k,t)/\nu$ and the collapse of the three curves indicates  
that $n_{k_0}(k,t)$ for $k>0$ vanishes as $1/L$. The same holds for $1-n_{k_0}(k,t)$ at $k<0$.

\begin{figure}[tb]
\includegraphics[width=.8\linewidth,clip]{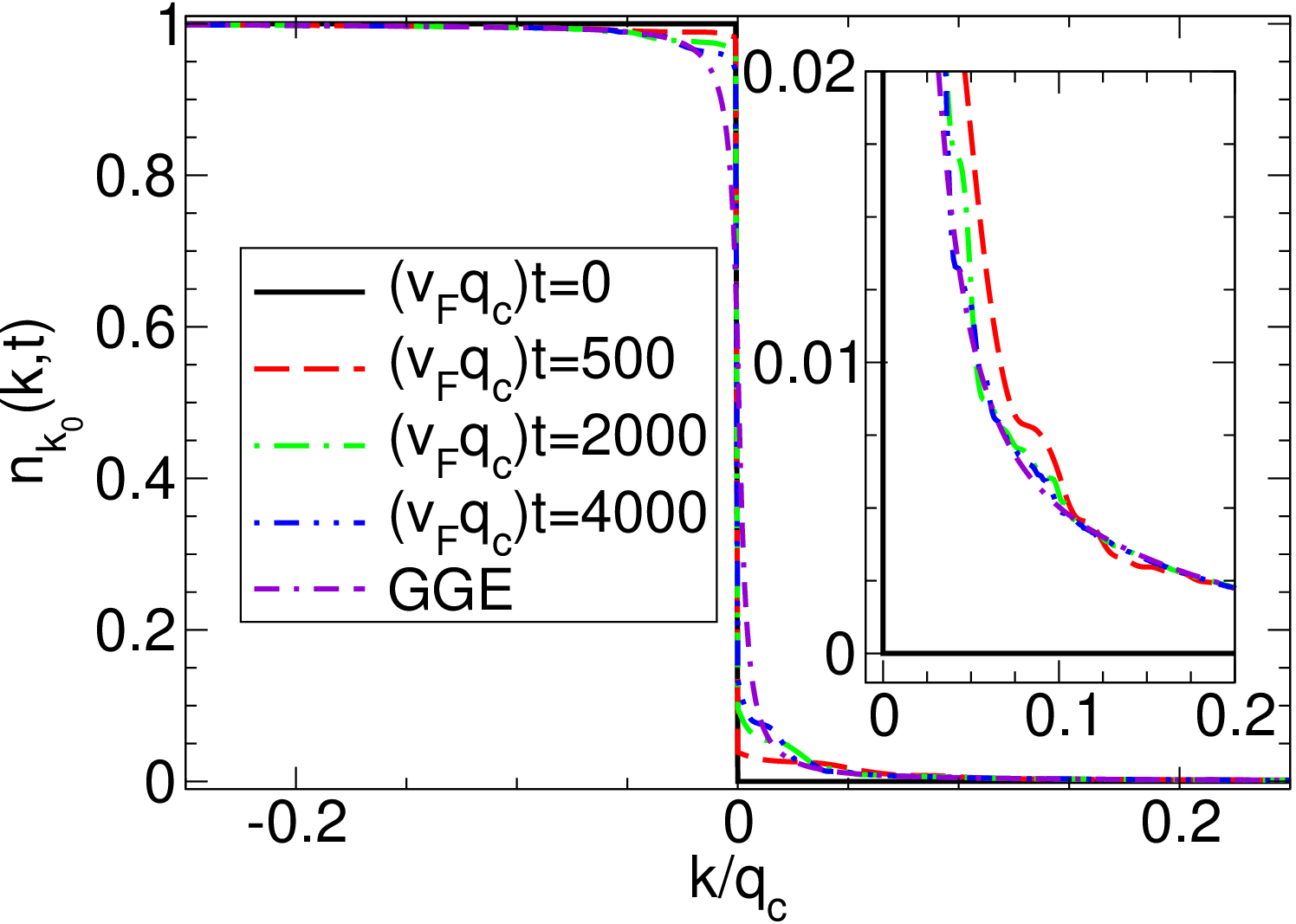}
\caption{(Color online) Momentum distribution function of the fermions $n_{k_0}(k,t)$ 
as a function of 
$k$ for different $t$. The GGE distribution obtained for the same system size is shown 
for comparison. The parameters are $\Gamma=0.03$, $\Omega=0.1$, $\nu=10^{-3}$,  
and $k_0/q_c=0.5$. }
\label{fig10}
\vspace{.2cm}
\includegraphics[width=.8\linewidth,clip]{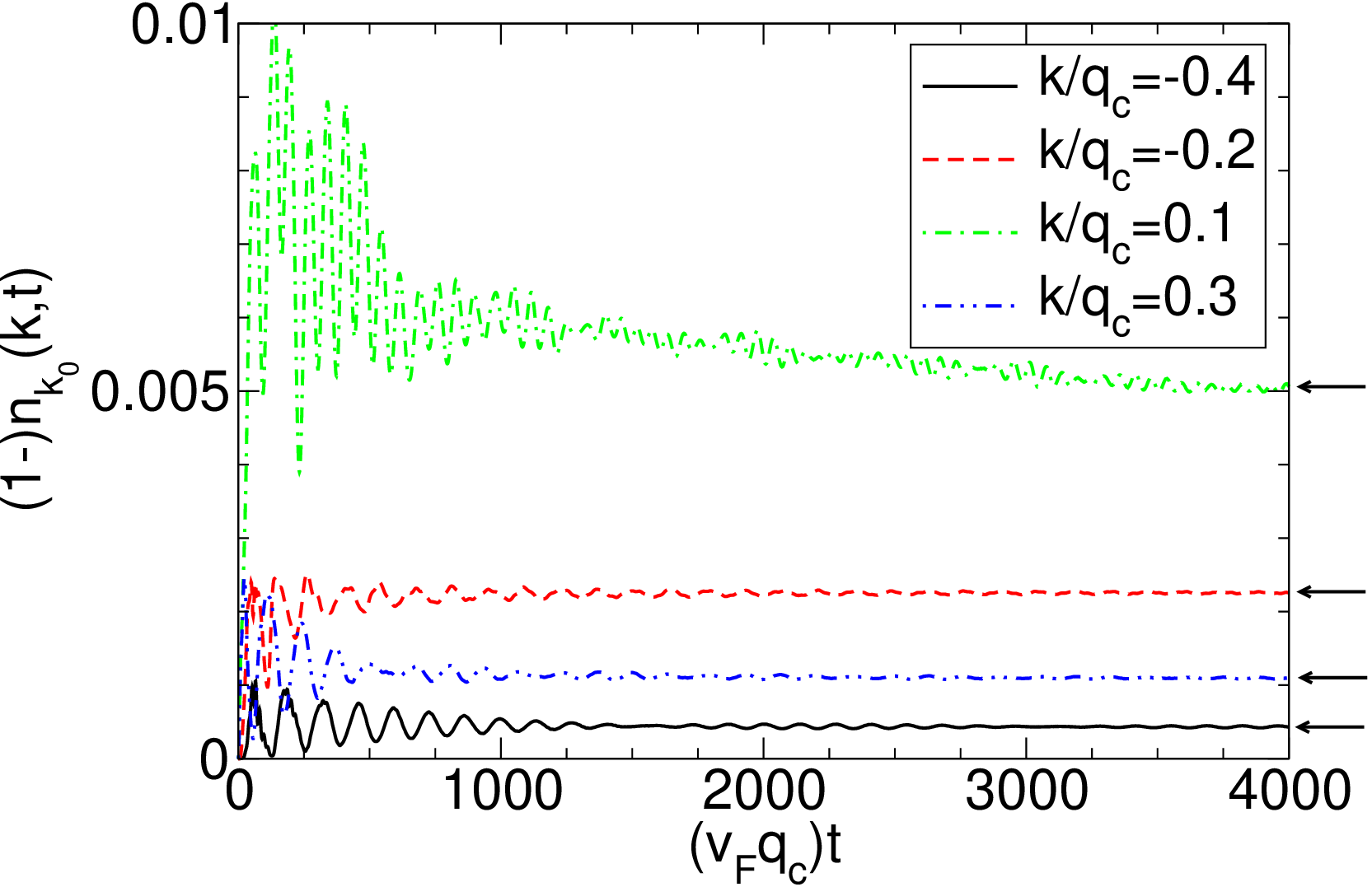}
\caption{(Color online) The same as in Fig.\ \ref{fig10}, but  
as a function of $t$ for different $k$. For the momenta 
$k<0$, $1-n_{k_0}(k,t)$ is shown. The GGE expectation values 
are indicated by the arrows.}
\label{fig11}
\end{figure}

After clarifying the behavior in the thermodynamic limit we now compare the time evolved 
and GGE distribution functions for fixed $L < \infty$. In Fig.\ \ref{fig10} we show the momentum 
distribution function as a function of $k$ for a few fixed $t$, the same model parameters as 
in Fig.\ \ref{fig5} (quench), and $k_0/q_c=0.5$. Up to very small momenta $|k| \ll q_c$ 
(see the inset) we find convergence towards the GGE prediction for times $t< t_{\rm r}$. 
As argued above the changes during the time evolution for the small momenta are large 
(from a step function at $t=0$ to a continuous function at large $t$) and convergence cannot 
be found for times $t < t_{\rm r}$. Note that because of the slow convergence the times shown 
in Fig.\ \ref{fig10} are much larger than those of Fig.\ \ref{fig5} even though the model parameters 
are the same in both figures. Finally, in Fig.\ \ref{fig11} we show $n_{k_0}(k,t)$ for a 
few fixed $k$ as a function of $t$. The GGE expectation values are indicated by the arrows.
For $k$ sufficiently far away from the initial step at $k=0$ it is again evident that 
$n_{k_0}(k,t)$ approaches $n_{\rm GGE}(k)$. In that sense a 
``quasi stationary state'' (finite $L$) is reached with a time averaged expectation 
value (over ``large'' times smaller than $t_{\rm r}$) which can accurately 
be described by the GGE prediction and which is different from the step function reached in the 
thermodynamic limit. In complete analogy to our earlier findings for the quench dynamics 
extending the time beyond $t_{\rm r}$ leads to enhanced oscillations with a time average which 
is still close to $n_{\rm GGE}(k)$. 

We note in passing that for the strongest possible electron-phonon coupling allowed by stability 
$\Gamma=\Omega$, $\lambda_-(q)=0$ for all $q<q_{\rm c}$. A similar situation is discussed in 
Ref.\ \onlinecite{Barthel}. In this case the amplitude of the oscillation in $n_{k_0}(k,t)$ for
fixed $k$ does {\it not} decay. The energies still approach a constant large time limit but 
with a $1/t$ instead of a $1/\sqrt{t}$ decay.\cite{Kennes}   

We did not succeed in extracting analytical results for the long-time asymptotics from the 
rather involved finite $L$ expression for $n_{k_0}(k,t)$ presented in the  
Appendix of  Ref.\ \onlinecite{VM1}. 
A numerical analysis for curves which at sufficiently large times (but still $t < t_{\rm r}$) 
oscillate around the finite $L$ GGE prediction (e.g. the $k/q_{\rm c}=-0.4$, $-0.2$, and $0.3$ 
curves in Fig. \ref{fig11}) shows that the amplitude decays {\it faster} than in the case of a quench, 
that is faster than $1/\sqrt{t}$. This has to be contrasted to the observation that it takes 
much {\it longer} times in case of the $k_0$-excitation before any asymptotic behavior sets 
in (compare Figs.\ \ref{fig6} and \ref{fig11}).  
From the numerical data it turned out to be impossible to extract an analytical form of 
the amplitude decay (e.g.\ power law $1/t^\chi$  with an exponent $\chi$ larger than 1/2). 
As already noted when discussing Figs.\ \ref{fig7} and \ref{fig8} the oscillation 
frequency of $n_{k_0}(k,t)$ at fixed $k$ seems to dependent on $k$. This becomes even more 
evident in Fig.\ \ref{fig11}. Besides this the oscillation frequency depends on the model 
parameters $\Gamma$ and $\Omega$. We did not succeed in extracting a clear picture for the 
dependence of the frequencies on these three parameters. In addition, it is not obvious how to 
relate the numerically determined frequencies appearing in $n_{k_0}(k,t)$ at fixed $k$ 
to the mode energies $\lambda_+(q)$ and $\lambda_-(q)$. This again exemplifies 
that the time evolution of the fermionic momentum distribution function for the 
$k_0$-excitation is rather involved and much richer than the dynamics encountered 
above.   

\section{Summary}
\label{sec:summary}

We studied the relaxation dynamics of an exactly solvable electron-phonon model out of 
two initial nonequilibrium states. This way we added a {\it continuum} model containing both 
{\it fermions and bosons} to the list of quantum many-body models recently investigated 
and considered other nonequilibrium situations than the parameter quench. Our model features 
a natural set of constants of motion, with as many elements as degrees of freedom. This set 
can be used to construct the initial state dependent generalized Gibbs ensemble.  We found that 
for {\it all observables} of interest to us the asymptotic long-time limit 
of expectation values which {\it become stationary} are equal to those obtained from 
the appropriate {\it natural GGE.} We discussed that the momentum distribution functions of 
the  GGE differ from those obtained within the canonical ensemble. While the (excess) energies 
and the electron momentum distribution function become stationary, the phonon momentum distribution 
at fixed momentum oscillates with a 
constant amplitude even at large times. This can be understood from the linear relation 
between the eigenmode ladder operators and the phononic ones. Long-time convergence of the energies 
and the fermionic momentum distribution function in the strict sense 
is only achieved if the thermodynamic limit is taken first. Going beyond this we showed that the 
GGE predictions for these expectation values agree to time averaged values taken at large times 
(even larger than the characteristic scale $t_{\rm r}=L/v_{\rm F}$). For most of the studied situations we 
were able to {\it analytically} describe how the GGE prediction is reached. We found 
that the asymptotic limit is only reached following a power law (instead of exponentially). 
The dynamics of the fermionic momentum distribution function resulting from the 
$k_0$-excitation turned out to be much richer than the one found for other observables and for the 
interaction quench.

\section*{Acknowledgments}
We are grateful to D.\ Schuricht, S.\ Manmana, M.\ Kollar, F.\ G\"ohmann, and A.\ Muramatsu 
for very useful discussions and to D. Schuricht for reading the manuscript prior to submission. 
Technical support by S.\ Grap and C.\ Karrasch is acknowledged. We thank  K. Sch\"onhammer 
for an earlier collaboration with one of us (VM) which is of fundamental importance for the 
present work.  We thank the Deutsche 
Forschungsgemeinschaft (Forscherguppe 912) for support.

\end{document}